\documentclass[10pt,journal,compsoc]{IEEEtran}
\ifCLASSOPTIONcompsoc
  \usepackage[nocompress]{cite}
\else
  \usepackage{cite}
\fi
\ifCLASSINFOpdf
  \usepackage[pdftex]{graphicx}
  \graphicspath{{./figures/}}
  \DeclareGraphicsExtensions{.pdf,.jpeg,.png,.pdf}
\else
  \usepackage[dvips]{graphicx}
  \graphicspath{{./figures/}}
  \DeclareGraphicsExtensions{.eps,.pdf}
\fi
\usepackage{url}

\usepackage{hhline}
\usepackage{enumitem}

\begin{document}
%
\title{Juxtaposing Controlled Empirical Studies in Visualization with Topic Developments in Psychology}
%
%
%
%

\author{Alfie~Abdul-Rahman,~\IEEEmembership{Member,~IEEE},
        Rita~Borgo,
        Min~Chen,~\IEEEmembership{Member,~IEEE},
        Darren~J.~Edwards,
        and~Brian~Fisher,~\IEEEmembership{Member,~IEEE}.
\IEEEcompsocitemizethanks{\IEEEcompsocthanksitem A. Abdul-Rahman is with the Department of Informatics, King's College London, UK. Email: alfie.abdulrahman@kcl.ac.uk.
\IEEEcompsocthanksitem R. Borgo is with the Department of Informatics, King's College London, UK. Email: rita.borgo@kcl.ac.uk. \IEEEcompsocthanksitem M. Chen is with the Department of Engineering Science, University of Oxford, UK. Email: min.chen@oerc.ox.ac.uk. \IEEEcompsocthanksitem D. J. Edwards is with the College of Human and Health Science, Swansea University, UK. Email: d.j.edwards@swansea.ac.uk. \IEEEcompsocthanksitem B. Fisher is with the School of Interactive Arts and Technology, Simon Fraser University, Canada. Email: bfisher@sfu.ca.}
\thanks{Manuscript received XX XX, XXXX; revised XX XX, XXXX.}}

%
%

\markboth{IEEE TRANSACTIONS ON VISUALIZATION AND COMPUTER GRAPHICS,~Vol.~XX, No.~X, September~2019}%
{Abdul-Rahman \MakeLowercase{\textit{et al.}}: Juxtaposing Controlled Empirical Studies in Visualization with Topic Developments in Psychology}
%



\IEEEtitleabstractindextext{%
\begin{abstract}
Empirical studies form an integral part of visualization research. Not only can they facilitate the evaluation of various designs, techniques, systems, and practices in visualization, but they can also enable the discovery of the causalities explaining why and how visualization works. This state-of-the-art report focuses on controlled and semi-controlled empirical studies conducted in laboratories and crowd-sourcing environments. In particular, the survey provides a taxonomic analysis of over 129 empirical studies in the visualization literature. It juxtaposes these studies with topic developments between 1978 and 2017 in psychology, where controlled empirical studies have played a predominant role in research. To help appreciate this broad context, the paper provides two case studies in detail, where specific visualization-related topics were examined in the discipline of psychology as well as the field of visualization. Following a brief discussion on some latest developments in psychology, it outlines challenges and opportunities in making new discoveries about visualization through empirical studies. 
\end{abstract}

}

\maketitle

\IEEEdisplaynontitleabstractindextext

%
\IEEEpeerreviewmaketitle

\IEEEraisesectionheading{\section{Introduction}\label{sec:introduction}}

%
%
%
%
\IEEEPARstart{E}{mpirical} studies play a significant role in the field of visualization \cite{Kosara:2003:CGA}. While they are often used to evaluate different visual designs, visualization techniques, and software systems \cite{Lam:2012:TVCG}, more and more studies were designed to gain fundamental understanding about why and how visualization works.

Empirical studies can take many forms, including controlled experiments, structured surveys and questionnaires, unstructured or free-text surveys, focus group discussions, and think aloud, case studies, field observation, laboratory observation, interviews, games, log analysis, algorithmic performance measurement, quality metrics, and so on. The survey by Lam et al.~\cite{Lam:2012:TVCG}, which focuses on the purpose of evaluation, provides a number of examples of the different typologies.  

A large number of empirical studies published as independent research papers in the visualization literature are in the form of controlled experiments, including controlled laboratory environments and semi-controlled crowd-sourcing environments. The reference section of this survey includes more than 129 references about these controlled and semi-controlled studies, providing a relatively comprehensive collection of these empirical studies. To our best knowledge, however, there are so far only two surveys on empirical studies in specific areas, namely glyph-based visualization \cite{Fuchs:2016:TVCG} and geo-spatial visualization (cartography) \cite{Roth:2017:IJC}. There is also a brief overview and categorization of controlled experiments in \cite{Kijmongkolchai:CGF:2017}.

Controlled experiments are the most predominant research methods in psychology, and there are hundreds of thousands of such studies in the psychology literature. In comparison, the controlled experiments in visualization are drops in the ocean. One would naturally be interested in how the controlled experiments in visualization relate to those in psychology. Have some phenomena in visualization already been well-studied in psychology? Does visualization present any new problems and hypotheses that are well worth the attention of both disciplines?

In this work, we conduct a comparative study to juxtapose the existing controlled empirical studies in visualization with topic developments in psychology. We aim to provide visualization researchers with a state-of-the-art report about such experiments in visualization and a temporal overview of the landscape in psychology. We aim to enable visualization researchers to relate the perceptual and cognitive phenomena in visualization to the existing developments in psychology through the use of visual analytics techniques, while informing psychology researchers about the unanswered perceptual and cognitive questions in visualization.

The main objective of this survey is therefore to fill in a major gap in the literature. Visualization literature features many contributions in the form of survey reviews. Between 2002 and 2017, \emph{Computer Graphics Forum} published nearly 40 state-of-the-art reports or survey papers on topics in visualization. In their 2017 survey of surveys, McNabb and Laramee selected and examined 86 surveys~\cite{McNabb:2017:CGF}. Among these, however, only a small number have focused on human factors. These include reviews of evaluation studies on specific topics, such as eye tracking~\cite{Kurzhals:2016:IV}, mobile devices~\cite{Blumenstein:2016:Beliv}, parallel coordinates~\cite{Heinrich:2013:STAR, Johansson:2016:TVCG}, and streaming data~\cite{Dasgupta:2017:CGF}. Perhaps the most significant survey on empirical study is the paper by Lam et al.~\cite{Lam:2012:TVCG} where the authors examined a large collection of evaluation studies and categorized them into seven scenarios.

Furthermore many empirical studies in visualization are of a discovery nature. For example,
Borgo et al. and Correll et al. discovered humans' capability of visual averaging in pixel-based visualization~\cite{Borgo:2010:TVCG} and time series visualization~\cite{Correll:2012:CHI};
Haroz and Whitney explored the capability limits of attention in visualization~\cite{Haroz:TVCG:2012};
Rensink and Baldridge, and Harrison et al. detected signals suggesting that humans' perception of correlation may correlate with Weber's law~\cite{Rensink:2010:CGF, Harrison:TVCG:2014};
Chung et al. studied the orderability of visual channels~\cite{Chung:2016:CGF}; and
Kijmongkolchai et al. measured the soft knowledge used in visualization~\cite{Kijmongkolchai:CGF:2017}.
In many ways, these discovery studies are similar to a huge volume of empirical studies in psychology, except that they focus on perceptual and cognitive phenomena in visualization.

In addition, while evaluation studies may inform us about which designs, techniques, systems, or work practices are more effective than others, they also indirectly feature questions and answers about perception and cognition. It is highly desirable to juxtapose both discovery and evaluation studies in visualization with the empirical studies in psychology.

In the remainder of this paper, we first briefly describe the methodology adopted to compile this survey in Section \ref{sec:StudyMethod}. This is followed by a summary view of the broad landscape of psychology, in Section \ref{sec:PsyTaxonomy}, through the construction of a high-level taxonomy of psychology. This allows us to illustrative the process of taxonomy construction, while identifying some variables that may be used in categorizing empirical studies in visualization.
We then consider in detail a taxonomy for empirical studies in visualization in Section \ref{sec:ESVTaxonomy}, where we discuss various variables for categorization and reason about the options for ordering these variables in defining a taxonomy.

In Section \ref{sec:TopicAnalysis} we give a brief overview about the history of the discipline, the commonly-used organization of subjects or themes in the discipline, the major schools of thoughts, and the popular research methods. This is followed by a topic analysis of two major journals in psychology and the development trend of over 30 keywords.
In Section \ref{sec:Juxtaposition} we juxtapose the empirical studies in visualization with the topic developments in psychology. We use visualization to highlight the synergy between the two disciplines, topics where visualization researchers may potentially find many existing studies, topics that have significant impact on visualization but require new studies to address the complexity of visualization tasks, and topics that demand substantial new efforts from both disciplines.

This is followed by two case studies in Section \ref{sec:CaseStudies}, where, for each case study, we juxtapose empirical studies published in psychology journals and visualization journals (also including journals in other domains). This juxtapositional analysis allows us to observe the similar and different characteristics of empirical studies across the two domains, and appreciate that visualization-related empirical studies can not only help define a significant application area of psychology, but also provide opportunities to answer fundamental questions in visualization and to develop new computing techniques for supporting empirical studies.

In Section \ref{sec:NewTrends}, we briefly describe several recent developments in psychology and discuss their relevance to visualization.  
Finally, in Section \ref{sec:Challenges}, we summarize the challenges and opportunities in conducting empirical studies in visualization. We point out the need for visualization researchers to be familiar with the landscape and historical developments in psychology as well as the need to stimulate new hypotheses and new experiments based on phenomena and tasks in visualization. We emphasize that the need for conducting such studies in the field of visualization as well as the need to collaborate with researchers in psychology.

\section{Study Method}
\label{sec:StudyMethod}
Ideally one would like to conduct a comprehensive and comparative survey of controlled empirical studies in visualization as well as in psychology. While it is feasible to conduct a traditional survey on controlled empirical studies in visualization, it would be an enormous challenge to attempt a traditional survey on those studies in psychology. There are hundreds of publication venues in psychology (e.g., 111 journals listed by Wikipedia). We therefore use two different approaches to survey the two disciplines respectively.

\noindent\textbf{STEP A.} For controlled experiments in visualization, we use ``close reading'' to study over 129 research papers published in the visualization literature. We perform a taxonomic analysis of these papers by examining their categories under different classification schemes and by comparing different options in defining a taxonomic hierarchy for organizing these papers. This step is to be reported in detail in Section \ref{sec:ESVTaxonomy}.

\noindent\textbf{STEP B.} For controlled experiments in psychology, we use ``distant reading'' to study the temporal evolution of topics in psychology. We first establish an initial list of topics by constructing a high-level taxonomy of psychology based on seven textbooks and some ten online resources. The psychology-trained co-authors lead the charting of the overall landscape of psychology and scrutinizing of the details of category labeling, while computer-science-trained co-authors lead a systematic approach of identifying variables for categorization and proposing the ordering of these variables. This is to be reported in Section \ref{sec:PsyTaxonomy}.

We then use text analysis to examine papers published between 1978 and 2017 (for 40 years) in two major journals of psychology: \emph{Behavioural and Brain Sciences} and \emph{Psychological Review}. We make use of software for topic analysis and visualization to handle a huge volume of data automatically. This algorithmic approach allows us to compile major statistical indicators about the topic developments in psychology. The psychology-trained co-authors in the team then analyze the results generated by the software, and make appropriate adjustments to the keywords and topics, which are used to rerun the algorithmic method. This combined human-machine process requires several iterations until the results offer a meaningful representation of the topic developments in psychology.
This is to be reported in Section \ref{sec:TopicAnalysis}.

\noindent\textbf{STEP C.} STEP A and STEP B, which are conducted in parallel, are followed by an integrated study in this third step. The controlled empirical studies in visualization surveyed in STEP A are first tagged with keywords and topics identified in STEP B. We then examine the relations between the controlled empirical studies in visualization and the topics in psychology using various visualizations where the entities of the two disciplines are juxtaposed and interconnected, reported in Section \ref{sec:Juxtaposition}. This facilitates further analysis of the challenges and opportunities in conducting empirical studies in visualization, reported in Section \ref{sec:Challenges}.

\begin{figure*}[t]
   \centering
   \includegraphics[width=\linewidth]{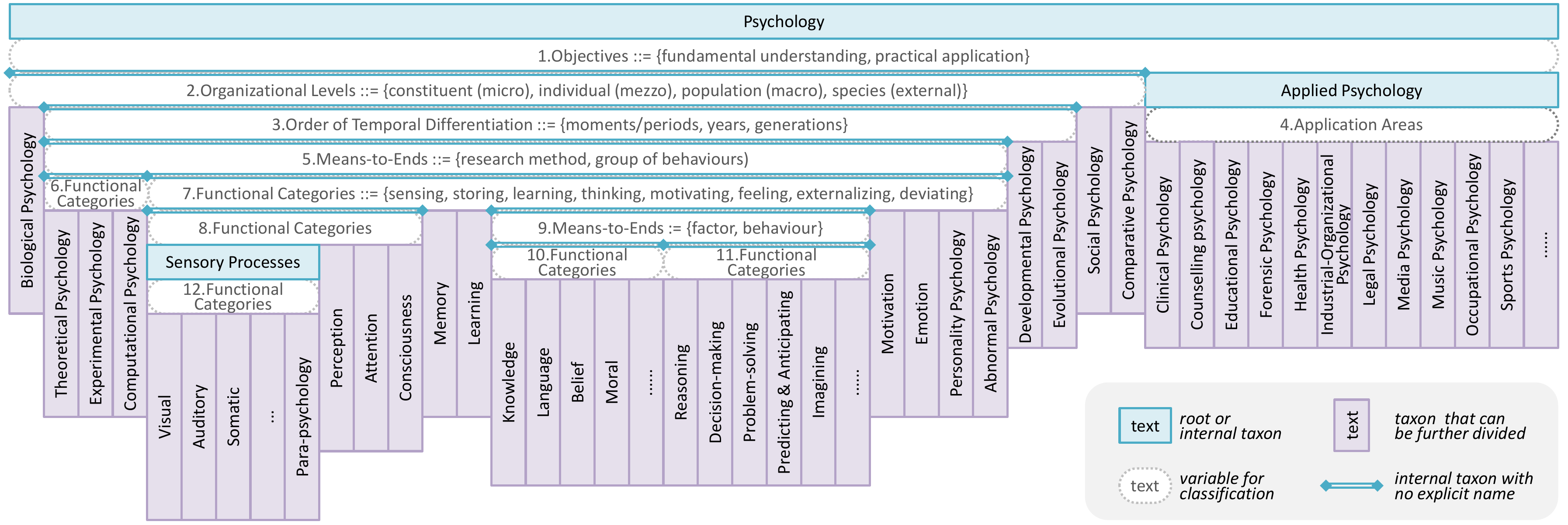}
   \caption{\label{fig:taxonomyH}
     A high level taxonomy of psychology, which is constructed based on a number of textbooks of psychology and a number of online resources on branches of psychology.}
   \vspace{-4mm}
\end{figure*}

\section{A High Level Taxonomy of Psychology}
\label{sec:PsyTaxonomy}
Psychology is a huge discipline. To our best knowledge there has not been any influential taxonomy proposed in literature to encompass most topics in psychology. One reason may be because of the sheer number of topics in the discipline. Another may be due to the difficulties for experts to reach a consensus. The latter reflects, to a large extent, the historical misunderstanding about the correctness of taxonomies.

In scientific and scholarly disciplines, a collection of concepts are commonly organized into a taxonomy, where concepts are known as ``taxa'' and are typically arranged hierarchically using a tree structure \cite{Chen:2017:CGA}. The main steps for building a taxonomy include:

\renewcommand{\labelenumi}{(\alph{enumi})}
\begin{enumerate}[noitemsep,nolistsep]
\item
mustering a collection of concepts (or entities);
\item
identifying a list of candidature variables, each of which is typically a nominal and ordinal variable and has a small number of valid values;
\item
using each variable to categorize the concepts (or entities) into groups and observe comparatively the distribution of the concepts (or entities) resulting from the application of the different variables;
\item
making the collection of all concepts (or entities) as the root of the taxonomic tree;
\item
selecting a principal variable and applying the selection to the current collection, where the valid values of the variable are thus the taxa;
\item
considering each group of concepts under a taxon as a new collection of concepts (or entities) and repeating steps (c-f) until there is no more than one concept under a taxon or when the taxonomic tree reaches a curtain depth.
\end{enumerate}

\noindent Step (b) and Step (e) are often the main sources of disagreement among experts as they involve subjective decisions as to what variables are to be considered and how these variables are selected or ordered. Any reasonable decision should yield a relatively useful taxonomy. There are many factors that may influence the judgment as to what can be considered a ``reasonable decision''. For example, one factor can be the desire to depict the commonly-used or commonly-accepted grouping in the upper part of a taxonomic tree. Another factor can be the preference for a more balanced tree, i.e., at each level of the tree, the taxa encompass similarly-sized collections of concepts. Some may be in favour of a variable with a smaller number of valid values, which usually leads to a deeper tree. Others may wish to have a shallower and flatter tree and may thus prefer to use multivariate variables or univariate variables that have relatively larger numbers of valid values. In many cases, there is no easy way to weigh different factors objectively, which often results in unnecessary discord. Therefore, we emphasize here that our proposed taxonomy is just one of many reasonable taxonomies that could be constructed. 

Figure~\ref{fig:taxonomyH} shows a high-level taxonomy of psychology constructed using the above process. We carefully assembled a collection of the major concepts based on chapter and section titles in seven textbooks in psychology \cite{Atkinson:1993:book,Smith:2003:book,Hockenbury:2010:book,%
Gleitman:2003:book,Cross:2005:book,Boff:1986:book1,Boff:1986:book2}, and some 10 lists of major branches of psychology, mostly on the web \cite{Wikipedia:2018:web,Cherry:2018:web1,Nordqvist:2018:web,Cherry:2018:web2,%
Sharma:2018:web,Revuu:2018:web,Netindustries:2018:web,HourGlass:2018:web,%
PsycholoGenie:2018:web,Ritchie:2009:book}. After considering a set of variables, we found that the first three variables as shown in Figure~\ref{fig:taxonomyH}  could be used to separate all branches of applied psychology as well as major branches such as \emph{Social Psychology}, \emph{Comparative Psychology}, \emph{Biological Psychology}, \emph{Developmental Psychology}, and \emph{Evolutionary Psychology}. These branches became taxa, each of which heads a sub-taxonomy. We then used the 5th variable ``Means-to-Ends'' to separate research methods from groups of behaviours to be studied.

For the latter, we introduced the 7th variable with eight functional categories, namely \emph{sensing}, \emph{storing}, \emph{learning}, \emph{thinking}, \emph{motivating}, \emph{feeling}, \emph{externalizing}, and \emph{deviating}. Among these, \emph{sensing} leads to a sub-taxonomy that includes major branches such as \emph{Sensory Processes}, \emph{Perception}, \emph{Attention}, and \emph{Consciousness}, while \emph{thinking} leads to a few major branches and significant topics in psychology. Meanwhile the other values of the 7th variable intrinsically correspond to other major branches such as \emph{Memory}, \emph{Learning}, \emph{Motivation}, \emph{Emotion}, \emph{Personality Psychology}, and \emph{Abnormal Psychology}.

While all taxa at the bottom of the taxonomic tree in Figure~\ref{fig:taxonomyH} can be further divided, this high-level taxonomy is adequate to be used a reference dimension for characterizing empirical studies in visualization in the next section.

\section{Taxonomy of Controlled Empirical Studies in Visualization}
\label{sec:ESVTaxonomy}
Building on the references collected by Lam et al. \cite{Lam:2012:TVCG}, Kijmongkolchai et al. \cite{Kijmongkolchai:CGF:2017}, Fuchs et al. \cite{Fuchs:2016:TVCG}, and Roth et al. \cite{Roth:2017:IJC}, we identified a total of 129 papers of controlled experiments published in visualization literature. We focused on those controlled experiments published as independent research papers, mainly because locating small controlled experiments that are components of design study papers or application papers is not a trivial undertaking. This collection of references naturally becomes the collection of entities as required by Step (a) in the process of building a taxonomy (Section \ref{sec:PsyTaxonomy}).

\subsection{Variables for Categorization}
\label{sec:Variables}
Naturally all variables that are shown in Figure~\ref{fig:taxonomyH} can potentially be used to categorize empirical studies in visualization. Here we list them formally as a subset of candidature variables:
\begin{itemize}[noitemsep,nolistsep]
\item
\emph{Study Objectives} $P_1 ::= \{$\emph{fundamental understanding}, \emph{practical application}$\}$.
\item
\emph{Organizational Levels} $P_2 ::= \{$\emph{constituent (micro)}, \emph{individual (mezzo)}, \emph{population (macro)}, \emph{species (external)}$\}$.
\item
\emph{Order of Temporal Differentiation} $P_3 ::= \{$\emph{moments/periods}, \emph{years}, \emph{generations}$\}$.
\item
\emph{Application Areas} $P_4 ::= \{ A_1, A_2, \ldots, A_k \}$. Most branches of applied psychology also application areas of visualization. In addition, there are other application areas of visualization, e.g., computational fluid dynamics, which are not specifically considered as branches in psychology.  
\item
\emph{Means-to-Ends of Psychology Studies} $P_5 ::= \{$\emph{research method}, \emph{group of behaviors}$\}$.
\item
\emph{Functional Categories of Research Methods} $P_6 ::= \{$\emph{fundamental understanding}, \emph{practical application}$\}$.
\item
\emph{Functional Categories of Behaviors} $P_7 ::= \{$\emph{sensing}, \emph{storing}, \emph{learning}, \emph{thinking}, \emph{motivating}, \emph{feeling}, \emph{externalizing}, \emph{deviating}$\}$
\item
\emph{Functional Categories of Sensing} $P_8 ::= \{$\emph{sensory processes}, \emph{perception}, \emph{attention}, \emph{consciousness}$\}$.
\item
\emph{Means-to-Ends of Thinking} $P_9 ::= \{$\emph{factor}, \emph{behavior}$\}$.
\item
\emph{Functional Categories of Factors in Thinking} $P_{10} ::= \{$\emph{knowledge}, \emph{language}, \emph{belief}, \emph{moral}, $\ldots \}$.
\item
\emph{Functional Categories of Thinking Behaviors} $P_{11} ::= \{$\emph{reasoning}, \emph{decision-making}, \emph{problem-solving}, \emph{predicting and anticipating}, \emph{imagining} $\ldots \}$.
\item
\emph{Functional Categories of Sensory Processes} $P_{12} ::= \{$\emph{visual}, \emph{audio}, \emph{somatic}, $\ldots$, \emph{para-psychological}$\}$.
\end{itemize}

\begin{figure*}[t]
   \centering
   \includegraphics[width=\linewidth]{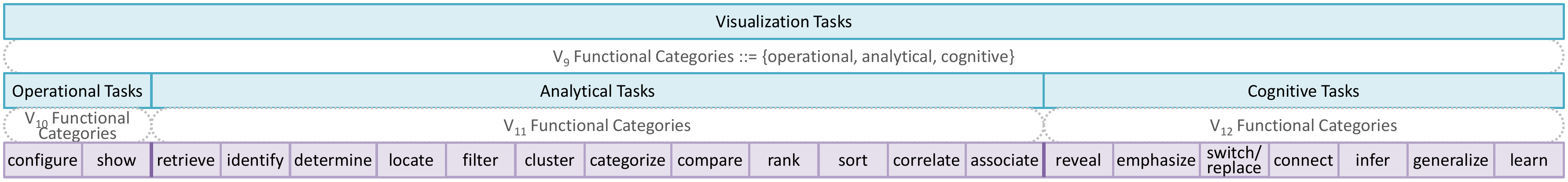}
   \caption{\label{fig:VisualizationTasks}
     A categorization scheme for characterizing different visualization tasks. Result of the integration of the task taxonomies proposed by  Wehrend and Lewis \cite{Wehrend:1990:Vis}, Zhou and Feiner \cite{Zhou:1998:CHI}, Amar et al. \cite{Amar:2005:InfoVis}, Valiati et al. \cite{Valiati:2006:BELIV}, and Pretorius et al. \cite{Pretorius:2014:LLCS}.}
   \vspace{-4mm}
\end{figure*}

Meanwhile, in the field of visualization, we often consider other variables that are not included in the above list. Some of these variables may be specific for visualization, while others may belong to the subtrees defined under some leaf taxa as in Figure~\ref{fig:taxonomyH}.

Lam et al. \cite{Lam:2012:TVCG} considered ``scenario'' as a variable that distinguishes a wide range of study methods, including controlled experiments, surveys and questionnaires, focus group discussions, and think aloud, case studies, field observation, laboratory observation, interviews, games, log analysis, algorithmic performance measurement, and quality metrics. This is indeed a multivariate variable that can be decomposed into several variables, such as:

\begin{itemize}
\item
\emph{Study Platforms} $V_1 ::= \{$\emph{laboratory}, \emph{internet}, \emph{in the wild}$\}$.
\item
\emph{Types of Intervention} $V_2 ::= \{$\emph{observation}, \emph{interview}, \emph{focus group discussion}, \emph{stimuli-and-responses}$\}$.
\item
\emph{Study Types} $V_3 ::= \{$\emph{exploratory studies}, \emph{assessment studies}, \emph{manipulation experiments}, \emph{observation experiments}$\}$ \cite{Cohen:1995:book}.
\item
\emph{Types of Collected Data} $V_4 ::= \{$\emph{free text}, \emph{structured survey results}, \emph{structured behavior data}, \emph{unstructured behavior data}, \emph{cognitive activity data}$\}$.
\end{itemize}

Kijmongkolchai et al. \cite{Kijmongkolchai:CGF:2017} considered two variables in their two-dimensional categorization. The first variable broadly divides all empirical studies into two groups, ($a_1$) to gain new understanding about perception and cognition in the context of visualization, and ($a_2$) to evaluate and compare visualization designs, algorithms, techniques, and systems. The second variable examines the different types of knowledge required to perform tasks featured in the studies, including ($b_1$) context, ($b_2$) pattern, and ($b_3$) statistics. The varying knowledge about contexts is expected to have significant impact when there are variations of the underlying data spaces, e.g., between different applications. This also includes context changes induced by algorithms and interactions, which result in changes to participants' attention to different parts of a data space. The varying knowledge about patterns is expected to have significant impact when there are changes of the visual representations of the same data. These studies are typically used to examine participants' performance in observing relatively complex information (e.g., features, patterns, events, etc.) rather than individual numbers. The varying ability to perceive statistical information is expected to have significant impact when there are changes to the visual representations related to specific numbers and statistical measures. These studies are typically used for examining participants' performance in determining the values of individual measures through visualization. These two variables are summarized below:

\begin{itemize}[noitemsep,nolistsep]
\item
\emph{Purposes of Studies} $V_5 ::= \{$\emph{fundamental understanding}, \emph{technical evaluation}$\}$.
\item
\emph{Scopes of Knowledge} $V_6 ::= \{$\emph{context}, \emph{pattern}, \emph{statistics}$\}$.
\end{itemize}

There are many other variables, for example:

\begin{itemize}[noitemsep,nolistsep]
\item
\emph{Design-Analysis Strategies} $V_7 ::= \{$\emph{between-subjects}, \emph{within-subjects}, \emph{mixed}, \emph{neither}$\}$.
\item
\emph{Metrics and Measures} $V_8 ::= \{$\emph{error}, \emph{time}, \emph{confidence}, \emph{attention}, \emph{motion}, \emph{spatial ability}, \emph{interpretation}, \emph{comprehension}, \emph{learning}, $\ldots \}$.
\end{itemize}

One important variable is the variation of visualization tasks, which has been studied extensively. A number of task taxonomies have been proposed (e.g., \cite{Buja:1996:JCGS,Pfitzner:2003:APSIV,Lammarsch:2012:EuroVA,Brehmer:2013:TVCG,Schulz:TVCG:2013,Ahn:2014:TVCG,Kerracher:2015:TVCG,Rind:2015:IV,Murray:2016:BioVis}). 
 For this survey, we make use of the proposals by Wehrend and Lewis \cite{Wehrend:1990:Vis}, Zhou and Feiner \cite{Zhou:1998:CHI}, Amar et al. \cite{Amar:2005:InfoVis}, Valiati et al. \cite{Valiati:2006:BELIV}, and Pretorius et al. \cite{Pretorius:2014:LLCS}. As illustrated in Figure \ref{fig:VisualizationTasks}, visualization tasks can be broadly divided into three categories, namely \emph{operational}, \emph{analytical}, and \emph{cognitive} tasks.
We found Andrienko and Andrienko's task taxonomy to provide a useful conceptual framework for differentiating between spatial and temporal data \cite{Andrienko:2006:book}.  The taxonomy features several levels and we could not find an easy way to integrate it with others.

\begin{itemize}[noitemsep,nolistsep]
\item
\emph{Task Groups} $V_9 ::= \{$\emph{operational}, \emph{analytical}, \emph{cognitive}$\}$.
\item
\emph{Functional Categories of Operational Tasks} $V_{10} ::= \{$\emph{configure}, \emph{show}$\}$.
\item
\emph{Functional Categories of Analytical Tasks} $V_{11} ::= \{$\emph{retrieve}, \emph{identify}, \emph{determine}, \emph{locate}, \emph{filter}, \emph{cluster}, \emph{categorize}, \emph{compare}, \emph{rank}, \emph{sort}, \emph{correlate}, \emph{associate}$\}$.
\item
\emph{Functional Categories of Cognitive Tasks} $V_{12} ::= \{$\emph{reveal}, \emph{emphasize}, \emph{switch/replace}, \emph{connect}, \emph{infer}, \emph{generalize} \emph{learn}$\}$.
\end{itemize}

\subsection{Selecting Variables}

After we have obtained a candidature list of variables, $P_1, P_2, \ldots, P_{12}, V_1, V_2, \ldots, V_{12}$, we analyze their usefulness for categorizing the controlled empirical studies that we have collected. It is obvious that all these empirical studies were carried out in the context of visualization, and most (if not all) do not fall into individual branches of applied psychology in Figure \ref{fig:taxonomyH}. Of course it is reasonable to suggest that these studies can be considered as applied psychology, and one may consider to create a new branch called \emph{Visualization Psychology}. Nevertheless, there is no reason to choose $P_1$ (\emph{Study Objectives}) and $P_4$ (\emph{Application Areas}) since it cannot divide the collection of studies further.

Similarly, hardly any empirical studies in visualization can be considered as part of \emph{Social Psychology} \emph{Comparative Psychology}, \emph{Biological Psychology}, \emph{Developmental Psychology}, and \emph{Evolutionary Psychology}, we can also eliminate $P_2$ (\emph{Organizational Levels}) and $P_3$ (\emph{Order of Temporal Differentiation}). Almost all the papers in the collection are not investigations into different research methods in psychology, there is also no reason to choose $P_5$ (\emph{Means-to-Ends of Psychology Studies}) and $P_6$ (\emph{Functional Categories of Research Methods}).

Because this survey focuses on controlled experiments, we do not expect any papers in the collection fall into the category \emph{in the wild} in terms of $V_1$ (\emph{Study Platforms}), \emph{observation}, \emph{interview} and \emph{focus group discussion} in terms of $V_2$ (\emph{Types of Intervention}), \emph{exploratory studies}, \emph{assessment studies}, and \emph{observation experiments} in terms of $V_3$ (\emph{Study Types}), and \emph{free text} and \emph{structured survey results} in terms of $V_4$ (\emph{Types of Collected Data}). On the other hand, $V_1$ can characterizes one of the main differences between controlled laboratory studies and semi-controlled crowd-sourcing studies, while $V_4$ can be used to distinguish those studies collecting structured behavior data (e.g., many typical accuracy-response studies) from those collecting unstructured behavior data (e.g., eye-tracking studies) and those collecting cognitive activity data (e.g., electroencephalography-based studies).

As shown in \cite{Kijmongkolchai:CGF:2017}, $V_5$ (\emph{Purposes of Studies}) and $V_6$ (\emph{Scopes of Knowledge}) can adequately divide the collection of controlled experiments into similarly-sized groups. We expect $V_7$ (\emph{Design-Analysis Strategies}) and $V_8$ (\emph{Metrics and Measures}) also have reasonable discrimination capacity.

The four variables for characterizing visualization tasks ($V_9$, $V_{10}$, $V_{11}$, and $V_{12}$) are no doubt important as specific visualization tasks are explicitly defined in most studies, especially those intended for technical evaluation. It is highly desirable to include these in a taxonomy for empirical studies in visualization. 

Tables \ref{tab:Categorization1}, \ref{tab:Categorization2}, and \ref{tab:Categorization3}  compare the categorizations using $V_5$ (\emph{Purposes of Studies}), $V_6$ (\emph{Scopes of Knowledge}), $V_{10} \lor V_{11} \lor V_{12}$ (\emph{Visualization Tasks}), and $P_7$ (\emph{Functional Categories of Behaviors}). These result from the Step (c) in the taxonomy construction process as described in Section \ref{sec:PsyTaxonomy}.

\begin{table*}[t]
  \centering
  \caption{A collection of visualization-related empirical studies in this survey categorized into $V_5$ (\emph{Purposes of Studies}), $V_6$ (\emph{Scopes of Knowledge}), $V_{10} \lor V_{11} \lor V_{12}$ (\emph{Visualization Tasks}), and $P_7$ (\emph{Functional Categories of Behaviors}).}
  \label{tab:Categorization1}
  \scalebox{1.0}{
  	\begin{tabular}{@{}l@{\hspace{2mm}}c@{\hspace{2mm}}c@{\hspace{2mm}}c@{\hspace{2mm}}c@{\hspace{2mm}}c@{}}
		\textbf{Paper} & $V_5$ \textbf{Purpose}  & $V_6$ \textbf{Knowledge} & $V_{10} \lor V_{11} \lor V_{12}$ \textbf{Visualization Tasks} & $P_7$ \textbf{Behaviors} \\
		\hline
		Adnan et al.~\cite{Adnan:2016:CHI} & Evaluation & Pattern, Statistics & Identify, Compare, Infer & Thinking \\
		Aigner et al.~\cite{Aigner:2011:CGF} & Evaluation & Statistics & Retrieve, Compare, Identify & Sensing \\
		Aigner et al.~\cite{Aigner:CGF:2012} & Evaluation & Pattern & 12 Tasks & Thinking \\
		Albers et al.~\cite{Albers:2014:CHI} & Evaluation & Statistics & Retrieve & Sensing \\
		Albo et al.~\cite{Albo:TVCG:2017} & Evaluation & Pattern, Statistics & 10 Tasks & Sensing, Thinking \\
		Alexander et al.~\cite{Alexander:2018:TVCG} & Both & Pattern, Statistics & Retrieve & Sensing \\
		Anderson et al.~\cite{Anderson:2011:CGF} & Understanding & Statistics & Compare & Sensing \\
		Bae \& Watson~\cite{Bae:TVCG:2014} & Understanding & Context, Pattern & Show & Learning \\
		Beecham et al.~\cite{Beecham:TVCG:2017} & Both & Pattern & Compare & Sensing \\	
		Bezerianos \& Isenberg~\cite{Bezerianos:TVCG:2012} & Understanding & Pattern & Retrieve & Sensing \\			
		Borgo et al.~\cite{Borgo:2010:TVCG} & Both & Pattern, Statistics & Retrieve, Determine, Compare & Sensing, Thinking \\
		Borgo et al.~\cite{Borgo:TVCG:2012} & Understanding & Context, Pattern, Statistics & Retrieve, Identify & Thinking, Storing \\
		Borgo et al.~\cite{Borgo:TVCG:2014} & Understanding & Pattern, Statistics & Retrieve & Sensing \\
		Borkin et al.~\cite{Borkin:TVCG:2013} & Understanding & Pattern & Retrieve, Identify & Storing \\
		Borkin et al.~\cite{Borkin:TVCG:2017} & Understanding & Context & Retrieve, Identify & Storing \\
		Boukhelifa et al.~\cite{Boukhelifa:TVCG:2012} & Evaluation & Pattern & Identify & Sensing \\
		Boy et al.~\cite{Boy:TVCG:2017}  & Evaluation & Context, Pattern & Infer & Thinking \\
		Boyandin et al.~\cite{Boyandin:CGF:2012} & Evaluation & Context, Pattern & Infer, Configure & Thinking \\
		Brandes et al.~\cite{Brandes:CGF:2013} & Understanding & Pattern & Compare, Determine & Thinking \\
		Bresciani \& Eppler~\cite{Bresciani:2009:TVCG} & Evaluation & Context & Configure, Determine & Thinking, Storing, \\
		 & & & & Externalizing \\
		Burch et al.~\cite{Burch:2011:TVCG} & Evaluation & Pattern, Statistics & Compare & Sensing, Thinking \\
		Cai et al.~\cite{Cai:2018:CGF} & Understanding & Statistics & Retrieve & Sensing \\
		Chen et al.~\cite{Chen:2006:TVCG} & Evaluation & Pattern & Determine & Thinking \\
		Chevalier et al.~\cite{Chevalier:TVCG:2014} & Evaluation & Pattern & Locate, Connect & Thinking \\
		Chung et al.~\cite{Chung:2016:CGF} & Understanding & Statistics & Sort, Rank & Thinking \\
		Cleveland \& McGill~\cite{Cleveland:1984:JASA} & Understanding & Statistics & Retrieve, Compare & Sensing \\
		Correll et al.~\cite{Correll:2012:CHI} & Understanding & Statistics & Retrieve, Compare & Sensing \\
		Correll et al.~\cite{Correll:2013:CHI} & Both & Pattern, Statistics & Retrieve & Sensing \\
		Correll \& Gleicher~\cite{Correll:TVCG:2014} & Evaluation & Pattern, Statistics & Identify, Infer & Thinking \\
		Correll \& Heer~\cite{Correll:2017:CHI} & Both & Statistics & Retrieve & Sensing \\
		Correll et al.~\cite{Correll:2019:TVCG} & Evaluation & Pattern, Statistics & Identify & Sensing \\
		Dasgupta et al.~\cite{Dasgupta:TVCG:2017} & Understanding & Context & Locate, Determine & Thinking \\
		Demiralp et al.~\cite{Demiralp:2014:TVCG} & Understanding & Statistics & Compare, Sort & Sensing \\
		Diehl et al.~\cite{Diehl:2010:TVCG} & Understanding & Pattern & Locate & Storing \\
		Dimara et al.~\cite{Dimara:2019:TVCG} & Evaluation & Pattern, Statistics & Retrieve & Sensing \\
		Dimara et al.~\cite{Dimara:TVCG:2017} & Understanding & Pattern & Compare & Motivating \\
		Etemadpour et al.~\cite{Etemadpour:TVCG:2015} & Evaluation & Context, Pattern, Statistics & Rank, Determine, Cluster & Thinking \\
		Felix et al.~\cite{Felix:2018:TVCG} & Understanding & Pattern, Statistics & Compare, Retrieve, Determine & Sensing, Thinking \\
		Fink et al.~\cite{Fink:TVCG:2013} & Understanding & Pattern & Retrieve & Sensing \\
		Fuchs et al.~\cite{Fuchs:TVCG:2014} & Evaluation & Pattern & Compare, Retrieve & Sensing, Thinking \\
		Ghani et al.~\cite{Ghani:2012:CGF} & Evaluation & Pattern & Locate & Storing \\
		Gleicher et al.~\cite{Gleicher:TVCG:2013} & Understanding & Statistics & Retrieve & Sensing \\
		Gramazio et al.~\cite{Gramazio:2017:TVCG} & Evaluation & Pattern, Statistics & Compare & Sensing \\
		Gramazio et al.~\cite{Gramazio:TVCG:2014} & Understanding & Pattern, Statistics & Identify & Sensing \\
		Griffin \& Robinson~\cite{Griffin:TVCG:2015} & Evaluation & Pattern & Locate, Connect, Associate & Thinking \\
		Gschwandtner et al.~\cite{Gschwandtner:TVCG:2017} & Evaluation & Pattern & Identify & Sensing \\
		Guo et al.~\cite{Guo:TVCG:2015} & Understanding & Pattern, Statistics & Identify, Compare & Thinking \\ \hline
	\end{tabular}
  }
\end{table*}


\begin{table*}[t]
  \centering
  \caption{A collection of visualization-related empirical studies in this survey categorized into $V_5$ (\emph{Purposes of Studies}), $V_6$ (\emph{Scopes of Knowledge}), $V_{10} \lor V_{11} \lor V_{12}$ (\emph{Visualization Tasks}), and $P_7$ (\emph{Functional Categories of Behaviors}).}
  \label{tab:Categorization2}
  \scalebox{1.0}{
  	\begin{tabular}{@{}l@{\hspace{2mm}}c@{\hspace{2mm}}c@{\hspace{2mm}}c@{\hspace{2mm}}c@{\hspace{2mm}}c@{}}
		\textbf{Paper} &  $V_5$ \textbf{Purpose}& $V_6$ \textbf{Knowledge} & $V_{10} \lor V_{11} \lor V_{12}$ \textbf{Visualization Tasks} & $P_7$ \textbf{Behaviors} \\
		\hline
		Haroz \& Whitney~\cite{Haroz:TVCG:2012} & Understanding & Pattern & Identify, Compare, Determine & Sensing, Thinking \\
		Haroz et al.~\cite{Haroz:2015:CHI} & Understanding & Pattern, Statistics & Retrieve & Storing \\
		Haroz et al.~\cite{Haroz:TVCG:2016} & Evaluation & Pattern & Infer & Thinking \\
		Harrison et al.~\cite{Harrison:TVCG:2014} & Understanding & Pattern, Statistics & Retrieve, Compare & Sensing \\
		Heer \& Bostock~\cite{Heer:2010:CHI} & Both & Statistics & Retrieve, Compare & Sensing \\
		Heer et al.~\cite{Heer:2009:CHI} & Both & Statistics & Retrieve & Sensing \\
		H\"{o}ferlin et al.~\cite{Hoferlin:TVCG:2012} & Evaluation & Context, Pattern & Locate & Thinking \\
		Hofmann et al.~\cite{Hofmann:TVCG:2012} & Evaluation & Pattern & Compare, Identify & Thinking \\
		Huron et al.~\cite{Huron:2014:TVCG} & Evaluation & Context & Configure & Thinking, Externalizing\\
		Isenberg et al.~\cite{Isenberg:2011:TVCG} & Evaluation & Statistics & Retrieve & Sensing\\
		Jakobsen \& Hornb{\ae}k~\cite{Jakobsen:TVCG:2013} & Understanding & Context & Locate, Connect, Infer & Thinking \\
		Jakobsen et al.~\cite{Jakobsen2:TVCG:2013} & Evaluation & Context & Retrieve, Locate, Connect & Thinking \\
		Jansen \& Hornb{\ae}k~\cite{Jansen:TVCG:2017} & Understanding & Statistics & Retrieve & Sensing \\
		Javed et al.~\cite{Javed:2010:TVCG} & Evaluation & Statistics & Retrieve, Compare & Sensing \\
		Kanjanabose et al.~\cite{Kanjanabose:CGF:2015} & Understanding & Pattern, Statistics & Retrieve, Identify, Cluster & Sensing, Thinking \\
		Kersten-Oertel et al.~\cite{Kersten-Oertel:2014:TVCG} & Evaluation & Pattern & Compare & Sensing \\
		Kijmongkolchai et al.~\cite{Kijmongkolchai:CGF:2017} & Both & Context, Pattern, Statistics & Determine, Infer & Storing, Thinking \\
		Kim et al.~\cite{Kim:TVCG:2012} & Evaluation & Pattern & Compare, Infer & Thinking \\
		Kim \& Heer~\cite{Kim:2018:CGF} & Evaluation & Context, Statistics & Compare, Retrieve & Sensing \\
		Kuang et al.~\cite{Kuang:CGF:2012} & Understanding & Pattern, Statistics & Retrieve & Sensing \\
		Kurzhals et al.~\cite{Kurzhals:CGF:2013} & Evaluation & Pattern & Identify & Sensing, Thinking \\
		Kwon et al.~\cite{Kwon:TVCG:2016} & Evaluation & Pattern & Identify, Connect, Compare & Thinking, Storing \\
		Laidlaw et al.~\cite{Laidlaw:2001:Vis} & Evaluation & Pattern & Identify, Determine & Thinking \\
		Li et al.~\cite{Li:2010:IV} & Understanding & Pattern, Statistics & Retrieve & Sensing \\
		Liccardi et al.~\cite{Liccardi:2016:CHI} & Understanding & Context, Pattern & Infer & Thinking \\
		Lin et al.~\cite{Lin:CGF:2013} & Understanding & Context & Compare, Infer & Thinking \\
		Lind \& Bruckner~\cite{Lind:2017:TVCG} & Evaluation & Context, Pattern & Identify, Compare & Sensing \\
		Livingston \& Decker~\cite{Livingston:2011:TVCG} & Evaluation & Pattern & Compare & Sensing \\
		Livingston et al.~\cite{Livingston:TVCG:2012} & Evaluation & Pattern & Determine & Thinking \\
		MacEachren et al.~\cite{MacEachren:TVCG:2012} & Evaluation & Pattern & Identify & Sensing \\
		Marriott et al.~\cite{Marriott:TVCG:2012} & Both & Pattern & Learn & Storing \\
		Mazurek \& Waldner~\cite{Mazurek:2018:CGF} & Evaluation & Context & Determine & Thinking \\
		Micallef et al.~\cite{Micallef:TVCG:2012} & Understanding & Context, Pattern & Retrieve, Infer & Storing, Thinking \\
		Mittelst\"{a}dt \& Keim~\cite{Mittelstadt:CGF:2015} & Understanding & Context & Retrieve & Sensing \\
		Morris et al.~\cite{Morris:2000:SPIE} & Understanding & Statistics & Retrieve & Sensing \\
		Netzel et al.~\cite{Netzel:TVCG:2014} & Evaluation & Pattern & Compare, Determine & Thinking \\
		Netzel et al.~\cite{Netzel:TVCG:2017} & Evaluation & Pattern & Locate & Thinking \\
		Nowell et al.~\cite{Nowell:2002:InfoVis} & Evaluation & Pattern & Identify & Sensing \\
		Ondov et al.~\cite{Ondov:2019:TVCG} & Evaluation & Pattern & Compare & Storing, Sensing \\
		Ottley et al.~\cite{Ottley:TVCG:2017} & Understanding & Context, Statistics & Infer & Thinking \\
		Padilla et al.~\cite{Padilla:TVCG:2017} & Evaluation & Pattern, Statistics & Locate, Identify, Compare & Thinking \\
		Pandey et al.~\cite{Pandey:2015:CHI} & Understanding & Pattern, Statistics & Compare & Sensing, Thinking \\
		Pandey et al.~\cite{Pandey:2016:CHI} & Understanding & Pattern & Compare, Cluster, Configure, & Thinking \\
		& & & Determine, Associate & \\
		Poupyrev et al.~\cite{Poupyrev:1998:CGF} & Evaluation & Pattern & Locate & Sensing, Externalizing \\
		Ragan et al.~\cite{Ragan:TVCG:2013} & Evaluation & Pattern & Identify, Determine & Thinking \\
		Rensink \& Baldridge~\cite{Rensink:2010:CGF} & Understanding & Statistics & Retrieve & Sensing \\
		Ryan et al~\cite{Ryan:2019:TVCG} & Understanding & Pattern, Statistics & Determine & Sensing \\ \hline
	\end{tabular}
  }
\end{table*}


\begin{table*}[t]
  \centering
  \caption{A collection of visualization-related empirical studies in this survey categorized into $V_5$ (\emph{Purposes of Studies}), $V_6$ (\emph{Scopes of Knowledge}), $V_{10} \lor V_{11} \lor V_{12}$ (\emph{Visualization Tasks}), and $P_7$ (\emph{Functional Categories of Behaviors}).}
  \label{tab:Categorization3}
  \scalebox{1.0}{
  	\begin{tabular}{@{}l@{\hspace{2mm}}c@{\hspace{2mm}}c@{\hspace{2mm}}c@{\hspace{2mm}}c@{\hspace{2mm}}c@{}}
		\textbf{Paper} &  $V_5$ \textbf{Purpose}& $V_6$ \textbf{Knowledge} & $V_{10} \lor V_{11} \lor V_{12}$ \textbf{Visualization Tasks} & $P_7$ \textbf{Behaviors} \\
		\hline
		Saket et al.~\cite{Saket:2018:TVCG} & Understanding & Context, Statistics & Identify, Determine, Correlate & Sensing, Thinking \\
		 & & & Retrieve, Compare, Filter, Rank & \\
		Saket et al.~\cite{Saket:TVCG:2014} & Evaluation & Pattern & Identify & Sensing \\
		Saket et al.~\cite{Saket:CGF:2015} & Evaluation & Pattern & Locate, Retrieve & Thinking, Storing \\
		Saket et al.~\cite{Saket:CGF:2016} & Both & Pattern & Locate & Sensing, Feeling \\
		Sarvghad et al.~\cite{Sarvghad:TVCG:2017} & Understanding & Context & Identify, Infer & Thinking \\
		Schloss et al.~\cite{Schloss:2019:TVCG} & Understanding & Pattern, Statistics & Retrieve & Sensing \\
		Sher et al.~\cite{Sher:2017:CGF} & Understanding & Pattern, Statistics & Retrieve & Sensing \\
		Skau et al.~\cite{Skau:CGF:2015} & Understanding & Pattern, Statistics & Retrieve & Sensing \\
		Skau \& Kosara~\cite{Skau:CGF:2016} & Evaluation & Pattern, Statistics & Retrieve & Sensing \\
		Song \& Szafir~\cite{Song:2019:TVCG} & Both & Pattern, Statistics & Determine & Sensing \\
		Srinivasan et al.~\cite{Srinivasan:2018:CHI} & Evaluation & Statistics & Compare, Identify & Sensing \\
		Strobelt et al.~\cite{Strobelt:TVCG:2017} & Evaluation & Pattern & Locate & Sensing \\
		Talbot et al.~\cite{Talbot:TVCG:2012} & Understanding & Statistics & Retrieve & Sensing \\
		Szafir~\cite{Szafir:2018:TVCG} & Understanding & Pattern & Retrieve & Sensing \\
		Szafir et al.~\cite{Szafir:2016:TVCG} & Understanding & Pattern & Compare & Sensing \\
		Talbot et al~\cite{Talbot:2012:TVCG} & Understanding & Statistics & Retrieve & Sensing \\
		Talbot et al.~\cite{Talbot:TVCG:2014} & Evaluation & Pattern, Statistics & Compare & Thinking \\
		Tanahashi et al.~\cite{Tanahashi:CGF:2016} & Evaluation & Context, Pattern & Learn & Learning \\
		Tory~\cite{Tory:2003:VIS} & Evaluation & Pattern & Identify, Locate & Thinking \\
		Vande Moere et al.~\cite{VandeMoere:TVCG:2012} & Understanding & Context, Pattern & Identify & Thinking \\
		Volante et al.~\cite{Volante:TVCG:2016} & Evaluation & Pattern & Determine & Feeling, Thinking \\
		Wagner Filho et al.~\cite{WagnerFilho:2018:CGF} & Evaluation & Statistics, Pattern & Locate, Compare, Retrieve & Sensing, Externalizing \\
		 & & & Identify, Filter & \\
		Walker et al.~\cite{Walker:TVCG:2017} & Evaluation & Context, Pattern & Locate, Compare & Thinking \\
		Wang et al.~\cite{Wang:2018:TVCG} & Evaluation & Statistics, Pattern & Compare & Sensing \\
		Ware~\cite{Ware:1988:CGA} & Understanding & Statistics & Retrieve & Sensing\\
		Wu et al.~\cite{Wu:TVCG:2017} & Evaluation & Context & Infer, Determine & Thinking \\
		Wun et al.~\cite{Wun:CGF:2016} & Evaluation & Pattern & Configure & Thinking \\
		Xu et al.~\cite{Xu:TVCG:2012} & Evaluation & Pattern & Identify & Sensing \\
		Yang et al.~\cite{Yang:TVCG:2017} & Evaluation & Context, Pattern, Statistics & Locate, Determine & Thinking \\
		Yang et al.~\cite{Yang:2018:CGF} & Evaluation & Statistics, Pattern & Compare, Retrieve & Sensing \\
		Yost \& North~\cite{Yost:2006:TVCG} & Understanding & Statistics, Pattern & Identify, Compare, Determine & Sensing \\
		Zhao et al~\cite{Zhao:2017:TVCG} & Evaluation & Statistics & Retrieve, Compare & Sensing \\
		Zhao et al.~\cite{Zhao:2015:CHI} & Evaluation & Context & Compare, Determine & Thinking \\
		Zhao et al.~\cite{Zhao:2019:TVCG} & Evaluation & Statistics, Pattern & Retrieve, Compare, Correlate & Sensing, Thinking \\
		Zheng et al.~\cite{Zheng:TVCG:2013} & Evaluation & Pattern & Determine & Sensing \\
		Ziemkiewicz et al.~\cite{Ziemkiewicz:TVCG:2013} & Both & Pattern & Locate & Thinking \\ \hline			
	\end{tabular}
  }
\end{table*}


From the two tables, we can obtain the following statistics about the numbers of occurrences of different category labels:

\begin{itemize}[noitemsep,nolistsep]
\item
$V_5$ \emph{Purposes of Studies}:
\emph{fundamental understanding} (62), \emph{technical evaluation} (79). 
\item
$V_6$ \emph{Scopes of Knowledge}:
\emph{context} (29), \emph{pattern} (93), \emph{statistics} (61).
\item
$V_{10} \lor V_{11} \lor V_{12}$ \emph{Visualization Tasks}:
\emph{configure} (5), \emph{show} (1); \emph{retrieve} (48), \emph{identify} (31), \emph{determine} (24), \emph{locate} (19),
\emph{filter} (2), \emph{cluster} (3), \emph{categorize} (0), \emph{compare} (45), \emph{rank} (3), \emph{sort} (2), \emph{correlate} (2), \emph{associate} (2);
\emph{reveal} (0), \emph{emphasize} (0), \emph{switch/replace} (0), \emph{connect} (5), \emph{infer} (15), \emph{generalize} (0), \emph{learn} (2).
\item
$P_7$ \emph{Functional Categories of Behaviors}:
 \emph{sensing} (72), \emph{storing} (13), \emph{learning} (2), \emph{thinking} (59), \emph{motivating} (1), \emph{feeling} (2), \emph{externalizing} (4), \emph{deviating} (0).
\end{itemize}

It is necessary to note that the above numbers of occurrences  should be considered as crude approximations because the assignment of various labels can be subjective. In particular, the difference between different visualization tasks can be quite subtle and their classification can thus be ambiguous and imprecise. For example, the actions of visually \emph{identify}, \emph{determine}, or \emph{locate} something can be quite similar, the actions of \emph{rank} and \emph{sort} can be highly related, and the action of \emph{infer} can easily involve many other actions such as \emph{filter} and \emph{associate}.
When we labeled each paper in the collection, we tried to focus on the tasks that were explicitly stated in the paper and avoid the introduction of additional labels that were not be intentionally investigated by the study concerns.

\begin{figure*}[t]
   \centering
   \includegraphics[width=\linewidth]{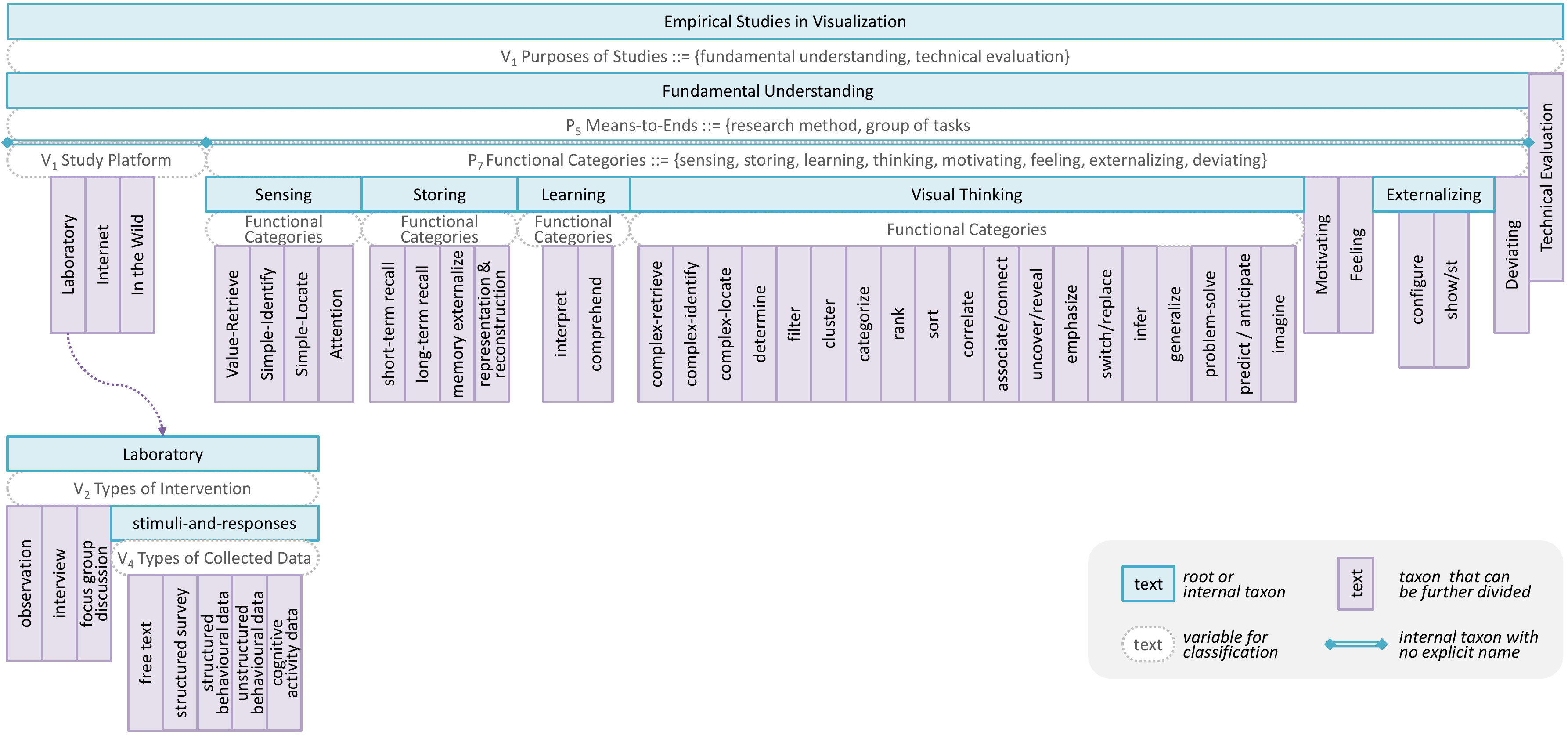}
   \caption{\label{fig:VisTaxonomy}
     A taxonomy for empirical studies in visualization, accommodating controlled, semi-controlled, and uncontrolled studies.}
     \vspace{-4mm}
\end{figure*}

\subsection{Examples of Classifying Empirical Studies}
In this subsection, we describe several examples to illustrate how we labeled each paper in the collection using some of the variables described in Section \ref{sec:Variables}. 
We adopted close reading to perform the labeling. 
Here we provide examples to illustrate our procedure.

The main objectives in the work by Borgo et al.~\cite{Borgo:2010:TVCG} is to explore effects of visual embellishments on memorization,
visual search and concept comprehension. The purpose ($V_5$) of the study was therefore classified as \emph{understanding}.
The study performed required participants to perform two tasks in parallel. The first corresponded to the primary task, e.g., stimuli exploration.
The primary task was subdivided into four main sections each probing a different aspect of the exploratory process with respect to memorization (long-term and working memory), visual search, and concept grasping hence \emph{retrieve} and \emph{identify} for visualization tasks ($V_{11}$ and $V_{12}$).
The stimuli used through the experiment were statistical representation of data as 2D histograms/bar charts and bubble charts in both numerical and metaphorical form. Participants were asked to identify and remember \emph{pattern}, \emph{statistics}, and \emph{context} hence the three categories chosen for knowledge ($V_6$).
The secondary task was performed in parallel with the primary acting as a distractor to mimic real life situations where focus of attention is continually challenged by the surrounding environment.
The study setting of two orthogonal tasks run in parallel explored user behavior ($P_7$) with respect to \emph{thinking} and \emph{storing} information in stressful situations.

 Chung et al.~\cite{Chung:2016:CGF} propose two empirical studies exploring human perception of orderability of the visual channels value, size, hue, texture, orientation, shape, and numerical representation. The purpose ($V_5$) of the study is therefore \emph{understanding}. The study analyzes orderability according to two main criteria:  perceived orderedness (e.g., sorting), value estimation e.g., if a target element has smallest value, largest value, or neither of the two (e.g., ranking). The two criteria were translated into corresponding tasks hence the choice of \emph{sort} and \emph{rank} for visualization tasks ($V_{11}$ and $V_{12}$).
Based on the nature of the tasks the behavior ($P_7$) was classified as \emph{thinking} while the knowledge measure ($V_6$) as \emph{statistics}. The latter is due to the estimation of a proportion as the cognitive aspect of the tasks.

Dimara et al.~\cite{Dimara:TVCG:2017} explore the ``attraction effect'' in visual layout. Attraction effect is defined as a cognitive bias in decision making when the choice between two alternatives is influenced by the presence of an irrelevant but stronger third alternative.
The purpose ($V_5$) of the study is again \emph{understanding} since the authors focus on the understanding of the nature of the bias.
A decision making task was at the core of the study where participants were asked to compare alternatives, which included both decoys and distractors, and choose the most appropriate one, this led to categorize visualization tasks ($V_{11}$) as \emph{compare}. The behavior ($P_7$) under study is that of \emph{motivating} since participant decisions were motivated by diverse appealing qualities of the various alternatives.
The authors controlled appeal of decoy and distractors following pre-defined patterns from decision-making research, they tried to determine if such patterns would lead to similar bias effects also in information visualization, we therefore categorized knowledge ($V_6$) as \emph{pattern}.

\begin {table*}[t]
 \caption {\label{tab:SchoolsOfThought}
 Commonly mentioned schools of thought in psychology \cite{Jastrow:1927:AJP,Byrnes:1992:EPR}.}
 \centering
 \scalebox{1.0}{
  \begin{tabular}{ l  c  l }
\textbf{School of Thought} & \textbf{Influential Figure} & \textbf{Main Belief or Emphasis}\\
\hline
Associationism & J. R. Angell (1869--1949) & Mental connections between events and ideas\\
Behaviourism & I. Pavlov (1849--1936) & Study of observable emitted behaviors\\
Cognitivism & J. Piaget (1896--1980) & Understanding how people think as state transitions\\
Constructivism & J. Dewey (1859--1952) & The mind actively gives meaning and order to the reality\\
Empiricism & J. Locke (1632--1704) & All knowledge is derived from experience\\
Functionalism & H. Ebbinghaus (1850--1909) & Mental operations and practical use of consciousness\\
Gestaltism & M. Wertheimer (1880-1943) & Study of holistic concepts, not merely as sums of parts\\
Nativism & I. Kant (1724--1804) & Certain skills/abilities are hard-wired into the brain at birth\\
Pragmatism & W. James (1842--1910) & Knowledge is validated by its usefulness\\
Structuralism & E. Titchener (1867--1927) & Analysis of consciousness into constituent components\\
\hline
 \end{tabular}
 }
 \vspace{-2mm}
\end{table*}

In Pandey et al.~\cite{Pandey:2016:CHI} the authors present a study exploring how human judge scatter plot similarity when scatter plots are presented as sets of icons. The study objectives was twofold: to understand the most dominant features of a scatter plot which influence human perception of similarity across scatter plots, and to measure the correlation between perceived similarity and exiting state of the art measures such as graph-theoretic scagnostics.
We therefore classified purpose ($V_5$) as \emph{understanding}.
The study main task asked participants to group scatter plots according to perceived similarity. Participants were presented with large collections of scatter plots and were free to attempt several grouping options before finalizing their answer.
The participant degree of freedom in interacting with the visualization and the type of perceptual judgment required by the task led us to classify the visualization tasks ($V_{11}$ and $V_{12}$) as \emph{compare}, \emph{cluster}, \emph{configure}, \emph{determine}, and \emph{associate}.
The nature of the task also implied that the measured participant behavior ($P_7$) was \emph{thinking}.
The study collected data reflected the patterns participants perceived within the scatter plots display therefore we classified knowledge ($V_6$) as \emph{pattern}.

The work of Tanahashi et al.~\cite{Tanahashi:CGF:2016} set itself aside. The authors in fact designed a study aimed at evaluating different design options employed in the creation of online guides and their effectiveness when applied to the design of online guides to educate novice users in the use of information visualizations. We therefore classified the study purpose ($V_5$) as \emph{evaluation}.
The authors measured the knowledge acquired by participants via comprehension tests, hence \emph{learning} as measured behavior ($P_7$) and \emph{learn} as visualization task ($V_{12}$). Comprehension tests required participants to answer multiple choice questions each contextual to the type of visualization and data shown.
Questions required the participants to interpret the visualization. We therefore categorized knowledge ($V_6$) as \emph{context} and \emph{pattern}.

\subsection{A Taxonomy of Empirical Studies in Visualization}

In order to build a general taxonomy for all empirical studies in visualization, it would be desirable to make use of $V_5$ (\emph{Purposes of Studies}) in a way similar to the high-level taxonomy of psychology in Figure \ref{fig:taxonomyH}.  It would also be desirable to use some combination of $V_1$ (\emph{Study Platforms}), $V_2$ (\emph{Types of Intervention}), $V_3$ (\emph{Study Types}), and $V_4$ (\emph{Types of Collected Data}) to separate, for instance, laboratory-based studies from crowd-sourcing studies, and eye-tracking studies from transcripts of focus group discussions.  It would also be highly desirable to include a sub-taxonomy for visualization, such as the one shown in Figure \ref{fig:VisualizationTasks}. Last but not least, it will be important to make connections with psychology by making use of a sub-taxonomy as in Figure \ref{fig:taxonomyH}.

Figure \ref{fig:VisTaxonomy} shows one possible option of such a taxonomy. As discussed in Section \ref{sec:PsyTaxonomy}, this is not necessarily the ``correct'' or ``best'' taxonomy, since such labeling is very unhelpful.
Because any appropriate selection of variables and reasonable ordering of the selected variables would likely result in a useful taxonomy, we believe that the taxonomy in Figure \ref{fig:VisTaxonomy} is such a reasonable taxonomy.

\section{Topic Developments in Psychology}
\label{sec:TopicAnalysis}
In this section, we first give a brief overview about the history of the discipline, the major schools of thoughts, and the popular research methods. We then describe technical process of conducting our survey of topic developments in psychology. Finally we present the results of a computer-assisted survey of two major journals in psychology between 1978 and 2017.

\subsection{Overview}
Psychology is the scientific study of behavior and mental processes. The earliest interest in humans' mind can be traced back to some 1500 years before common era (BCE). The development of this interest has continued ever since.    

The 16th century saw the beginning of western psychology, and attracted many great thinkers and practitioners at that time and the following centuries, such as
German philosopher Rudolf G\"{o}ckel (1547--1628),
French philosopher, mathematician, and scientist Ren\'{e} Descartes (1596--1650),
English doctor Thomas Willis (1621--1675),
English philosopher and physician John Locke (1632--1704),
Irish philosopher George Berkeley (1685--1753),
Scottish philosopher, historian, and economist David Hume (1711--1776), and many more \cite{Leahey:2017:book,Kardas:2013:book}.

The 19th century saw the emergence of psychology as an independent discipline rising from a branch of philosophy. Today, psychology is one of most popular academic disciplines, and is studied in the majority of universities around the world. Figure~\ref{fig:taxonomyH} outlines the broad landscape of contemporary psychology.

Perhaps because of the heritage of philosophy, there have been many schools of thought in psychology \cite{Jastrow:1927:AJP,Byrnes:1992:EPR}. The rise and fall of these schools of thought often signifies the paradigm shift in the discipline. Table \ref{tab:SchoolsOfThought} summarizes some major schools of thought.

In psychology, there are different forms of psychological enquiries, such as controlled experiments, correlational studies, naturalistic observation, case studies, interviews, discourse analysis, and personal reflections \cite{Hockenbury:2010:book}. There have been many proposed conceptual models (often referred to as theories), which were usually informed by empirical studies and many empirical studies were designed to test such models. There have also been many attempts to describe the causal relations in human behaviors and cognition using computational models.

\begin{figure*}[t]
   \centering
   \includegraphics[width=\linewidth]{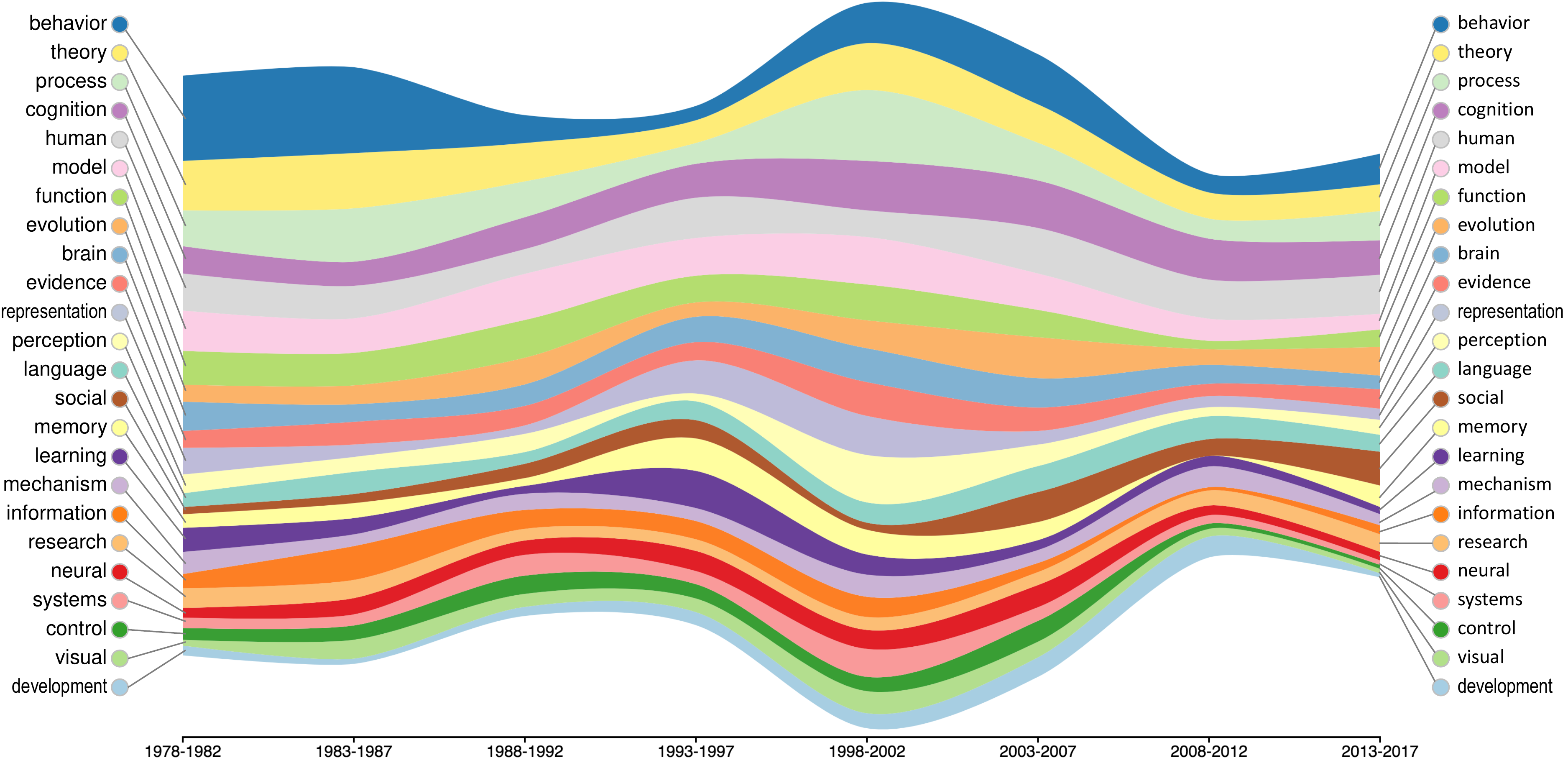}
   \caption{\label{fig:BBSTopics}
    Major keywords in Behavioural and Brain Sciences between 1978 and 2017.  The number of keywords are countered in blocks of five years to increase the number of papers being sampled at each sampling point. The ThemeRiver software interpolates the thickness between each pair of consecutive data points.}
    \vspace{-3mm}
\end{figure*}

\subsection{The Process of Topic Analysis}
In this work, we used three different types of visualization to study the topic development in psychology, namely tag clouds, temporal tag clouds, and ThemeRiver. We used Voyant Tools (\url{https://voyant-tools.org/}) for the text analysis. Voyant Tools is a web-based application that allows us to upload, process, generate data for text analysis (text mining tool) and visualization (uploads the file either in pdfs or text format). Voyant Tools process the text and automatically generates tag clouds and graphs, and  coverts data in downloadable csv or json format. Inside Voyant Tools, common stop words, such as the, an, etc., are automatically filtered. The Voyant Tools is widely used in the Digital Humanities field, and has a wide user base (\url{http://hermeneuti.ca/VoyantFacts}).

In the first phrase, we identified a number of high impact journals in psychology. We chose three journals, \emph{Annual Reviews of Psychology}, \emph{Perception}, and \emph{Vision Research}, to test the effectiveness of different parts of text in journal papers. For each of the three journals, we selected five articles in the same period. We carried out the text analysis for (i) full PDFs and (ii) Title+Abstract+Keywords (T-A-K), and generated a tag cloud for each journal (i.e., each group of five papers).
The psychology-trained co-authors reviewed and discussed the generated visualization, identified the need and method for filtering and combining words. We also found that it was not easy to compare different tag clouds, though comparing words within a tag cloud was effective.

After implementing the suggestions from psychology-trained co-authors in text analysis process, we were be able to generate informative tag clouds with minimal noise.  We found that using texts from the Title+Abstract+Keywords is better than full PDFs in most cases to provide us with visualizations that depict the main topics in each group of five papers.

In the second phase, we focused on two journals, \emph{Behavioral and Brain Sciences} and \emph{Psychological Review}. For the former, we only extract papers that are labeled as Target, Research, or Main articles. For the latter, we included all articles except errata.
As \emph{Psychological Review} does not have keywords for many articles, for consistency, we extracted Title+Abstract and Authors for articles in both journals.
We processed all issues of the two journals in the 40-year period of 1978--2017.

Using the Voyant Tools, we were able to generate a list of most popular terms from the Title+Abstract. These were then downloaded and saved as spreadsheets. The psychology-trained co-authors inspected these spreadsheets carefully, highlighted those terms that are more relevant to the field of psychology, and suggested words to be grouped together.

We focused the highlighted terms and grouped together words with similar semantics (i.e., plural and singular, British spelling and American spelling, present participles, and so on). 
using the refined data, we tested temporal tag clouds and ThemeRiver visualizations. We found that the ThemeRiver visualization is more effective in depicting the topic development over a period.
The temporal tag clouds and ThemeRiver were generated using D3.js. 


\begin{figure*}[t]
   \centering
   \includegraphics[width=\linewidth]{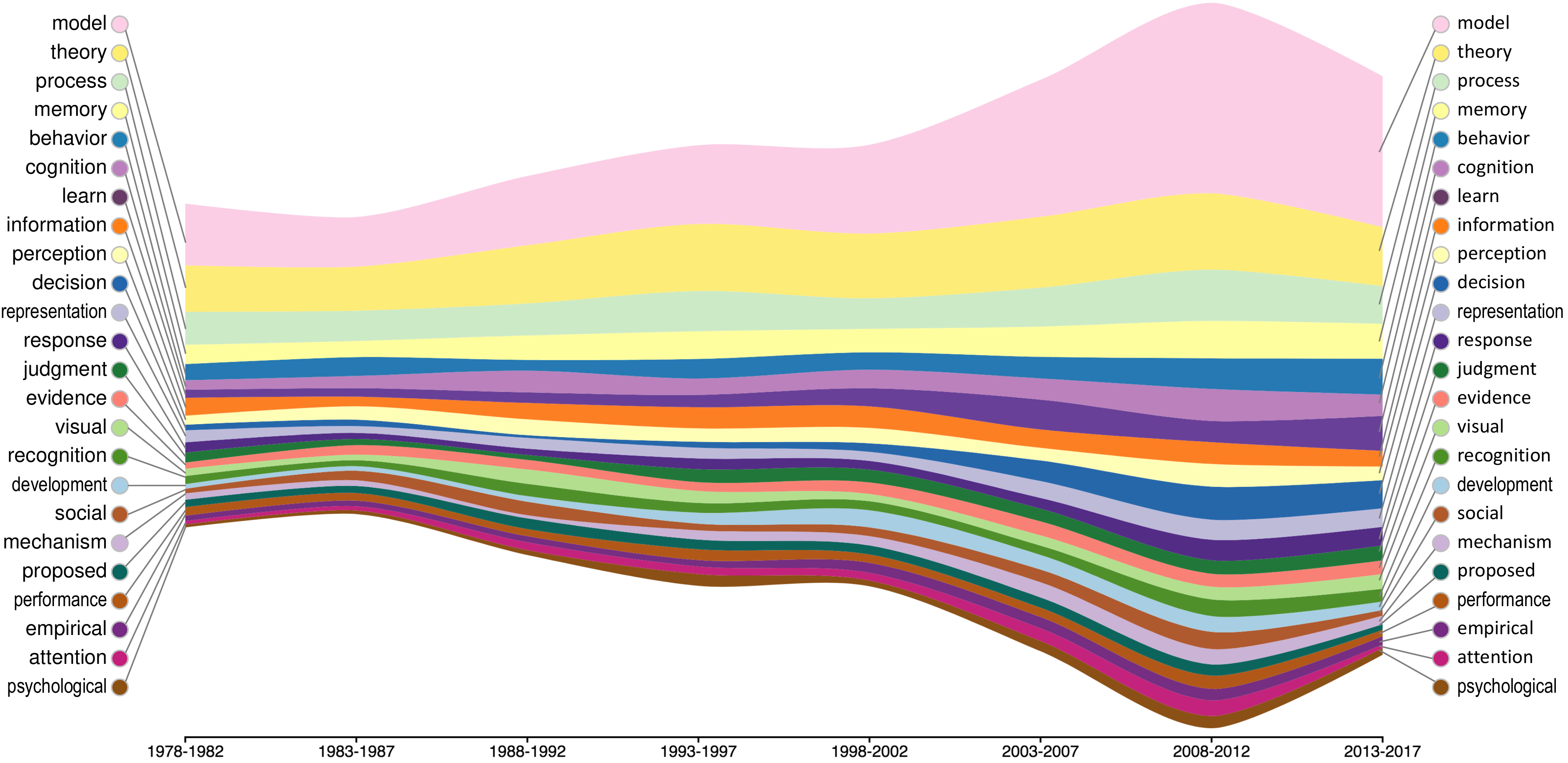}
   \caption{\label{fig:PRTopics}
    Major keywords in \emph{Psychological Review} between 1978 and 2017. The number of keywords are countered in blocks of five years to increase the number of papers being sampled at each sampling point. The ThemeRiver software interpolates the thickness between each pair of consecutive data points.}
    \vspace{-3mm}
\end{figure*}

\subsection{Topics in Psychology between 1978 - 2017}
Figure \ref{fig:BBSTopics} shows a ThemeRiver visualization that depicts the changes of the top 25 keywords in Behavioural and Brain Sciences between 1978 and 2017. The number of keywords are countered in blocks of five years to increase the number of papers being sampled at each sampling point. This results in 8 data points per keyword for 1978--1982, 1983--1987, $\ldots$, 2008-2012, 2013--2017. The ThemeRiver software interpolates the thickness between each pair of consecutive data points. From this visualization, we can make the following observations:

\begin{itemize}[noitemsep,nolistsep]
\item
In this journal, many uses of the term \emph{behavior} are usually associated with actual and specific behaviors, for which some behavioral outcomes are observed or detected under some conditions being studied. The the use of the term does not necessarily suggest behaviorism (conditioning), but it may be used in the context of behaviorism sometimes.
The steady decline in the use of this term may be due to a number of factors. For example, animal studies (e.g., rat studies) has been increasingly recognized as unethical, leading to fewer studies about animal behaviors. Another reason is that many authors have chosen to replace this term with more generally-accepted terms, such as ``cognition'', ``decision  making'', ``judgments'', and so on,  which give a less emphasis on the behavior itself. 
\item
There is a decline in using the term \emph{theory}. This is probably due to the fact that this journal has been gradually given less emphasis on general theory building and more focus on general review.
\item
The term \emph{model} also exhibits a decline. This is likely because of the same reason as for the term \emph{theory}.
\item
For the term \emph{perception}, there seems to be a spike in the period of 1998--2002, possibly due to the popularity of the subject area at this time, especially in areas of \emph{Applied Psychology}.  Note that there is a spike in the total number of papers in the journal during that period (about 25\% more than the previous and following half decades). Nevertheless the spike is still significant even after taking this fact into account.
\item
The term \emph{memory} sees a surge in the 15 year period of 1993--2007. This is likely because of the new emphasis on cognition during that period, perhaps together with the rising popularity of cognitivism. 
\item
The terms \emph{neural} and \emph{brain} typically represent the emerging trend in \emph{Biological Psychology}. The lack of an increasing pattern in both cases is somehow unexpected.
\item
The term \emph{learning} is prominent in the period of 1993--1997. Its decreasing pattern after that period may be related to the decreasing trend in using the term \emph{behavior}, since conceptually there are some overlaps among ``behavior'', ``behaviorism'', and ``learning''. In different periods of time, they may be sometimes more and other times less popular in use.
\end{itemize}

Figure \ref{fig:PRTopics} shows a ThemeRiver visualization that depicts the changes of the top 25 keywords in Psychological Review between 1978 and 2017. In the same way as for Behavioural and Brain Sciences, the number of keywords are countered in blocks of five years to increase the number of papers being sampled at each sampling point. The ThemeRiver software interpolates the thickness between each pair of consecutive data points. From this visualization, we can make the following observations:

\begin{itemize}
\item
The journal published noticeably more papers during the period between 2004 and 2010. Although the total number of papers (i.e., 250) published during the period of 2003--2007 is slightly more than that (i.e., 242) during 2008--2012, interestingly the spike occurs at the axial point of 2008--2012.
Most terms, such as \emph{memory}, \emph{attention}, \emph{social}, and \emph{representation} exhibit a gradual increasing trend in line with the growth of published papers in the journal.
\item
The term \emph{model} is overwhelmingly the most popular term in the journal, because there has been a focus on theory development in this journal.
There is also a large spike in the use of the term \emph{model} for the period of 2008--2012, possibly reflecting the journal's emphasis on model building during the period. 
\item
The uses of the terms \emph{decision} and \emph{learn} have slightly different trends as other terms with more growth in recent years. This may be attributed to the fact that they are often used in places where ``behavior'' or ``behavioral'' were used previously. The term \emph{decision} is often used in studies about \emph{memory}, \emph{attention}, \emph{categorization}, and so on.
\item
The term \emph{behavior} does appear to have an increasing trend in recent years. This is likely used to discuss actual behaviors in various tasks, instead of representing the paradigm of behaviorism. 
\end{itemize}

Because the journal of Behavioural and Brain Sciences has more general review papers on many different subject areas in psychology, it does not feature a highly dominant theme. Meanwhile the journal of Psychological Review has a greater 
 emphasis on developing models. This explains why the term \emph{model} is the most prominent.
The journal also gives an emphasis on \emph{theory} as the development of a model is typically informed by the development of a theory, while the testing of a model can provide evidence to support or falsify a theory.  
The term \emph{process} appears a lot in both figures, because it is a general word for scenarios where many cognitive components work together to process information in a way that can then be usable. This is in line with model building.  
The term \emph{memory} is also central to such processes as many models have a component of memory in order to process information. This suggest that the journal focuses largely on the paradigm of cognitivism, which is the dominant school of thought in psychology today.  
The term \emph{cognition} may appears less frequently than \emph{behavior}. This is largely because authors often use more specific words, such as ``visual'', ``recognition'', ``decision'', and ``response'',  instead of ``cognition''.

\section{Juxtapositional Analysis of Taxonomies and Topics}
\label{sec:Juxtaposition}
\begin{figure*}[t!]
	\centering
	\scalebox{1.0}{
	\begin{tabular}{@{}l@{}}
    \includegraphics[width=\linewidth]{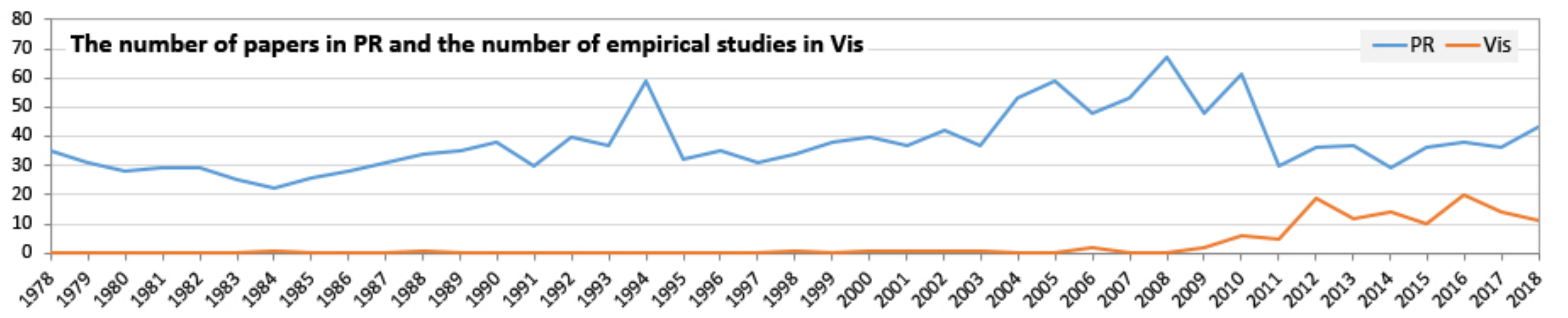}\\
  	{\small (a) Time series representing the numbers of papers in Psychological review and the number of papers on visualization-related}\\
  	{\small empirical studies in this survey.}\\
  	\includegraphics[width=\linewidth]{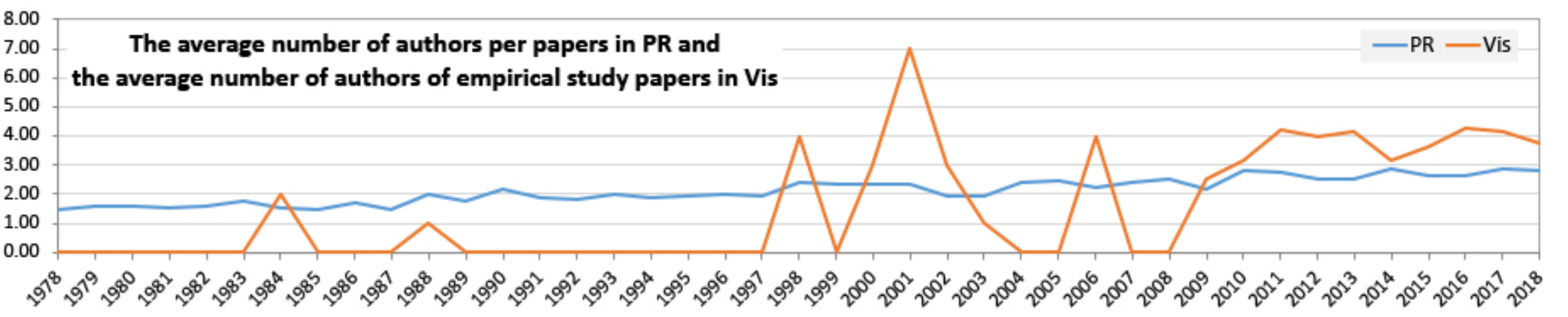}\\
  	{\small (b) Time series showing the average number of authors per paper in Psychological Review and the average number of authors of}\\
  	{\small empirical studies in visualization.}
  \end{tabular}
  }
  \caption{\label{fig:PRvsVis} Time series between 1978 and 2018 showing (a) the numbers of papers in Psychological Review (PR) and the number of papers on visualization-related empirical studies (Vis) collected in this survey; (b) the average of number of authors per paper in Psychological Review (PR) and average number of authors per paper for visualization-related empirical studies (Vis).}
  \vspace{-4mm}
\end{figure*}

Figure \ref{fig:PRvsVis}(a) shows two time series, representing the numbers of papers in Psychological Review between 1978 and 2018, and the number of papers on visualization-related empirical studies collected in this survey. Although these numbers are not directly comparable, we can make several inferred observations after taking some other factors into account.
\begin{itemize}[noitemsep,nolistsep]
    \item Although the collection of papers in Tables \ref{tab:Categorization1}, \ref{tab:Categorization2}, and \ref{tab:Categorization3} may not be complete, it is a relatively comprehensive collection, and thus the time series for the number of visualization-related empirical studies is indicative. From the orange-colored time series in Figure \ref{fig:PRvsVis}(a), we can observe that there were noticeably more papers since 2010 in comparison with the period before. Note that IEEE Visualization started in 1990, and EuroVis started in 1999. This suggests a healthy growth of empirical studies conducted in the context of visualization.
    \item From Figures \ref{fig:BBSTopics} and \ref{fig:PRTopics}, we have already observed the different trends of keyword counts in Behavioural and Brain Science and Psychological Review. In fact, the two journals also have different trends in terms of number of papers. We thus do not treat the variations exhibited by the blue-colored time series as a representative trend in psychology. It is also necessary to point out that not all papers in Psychological Review report empirical studies. Nevertheless, most papers in this journal report empirical studies and/or models derived or based on the results of empirical studies. Hence it is reasonable to consider that this journal alone has published more empirical studies than the field of visualization. Considering that there are over 100 journals in the field of psychology, the number of empirical studies published in visualization venues are drops in the ocean. From the discussions on taxonomies in Sections \ref{sec:PsyTaxonomy} and \ref{sec:ESVTaxonomy}, the subject of visualization no doubt shares much common ground with the discipline of psychology. Meanwhile, through this survey, we have also observed that there have not been many visualization-related empirical studies published in psychology journal. Hence, there is clearly a need and scope for more visualization-related empirical studies.
    \item In addition, from Figure \ref{fig:PRvsVis}(b) we can observe that the average number of authors per paper in Psychological Review exhibits a slow trend of gradual increase between 1978 and 2018. In comparison, the average number of authors per paper for visualization-related empirical studies is in general higher during the period between 2010 and 2018. Note that the averages between 1978 and 2009 are statistically not meaningful and should not be considered in the comparison.
\end{itemize}

With the high-level taxonomies developed in Sections \ref{sec:PsyTaxonomy} and \ref{sec:ESVTaxonomy}, we can examine the relations between the topics in psychology and empirical studies in visualization. In particular, we can use the variable $P_7$, \emph{Functional Categories of Behaviors}, to categorize the collection of empirical studies as shown in Tables \ref{tab:Categorization1}, \ref{tab:Categorization2}, and \ref{tab:Categorization3}.

We first notice that many papers are about the behaviors of \emph{sensing} (72) and \emph{thinking} (59). In many ways, this is expected. However, there are also 13 papers about \emph{storing}, 4 about \emph{externalizing}, 2 about \emph{learning}, 2 about \emph{feeling}, and 1 about \emph{motivating}. There is none about \emph{deviating}. In comparison with the ThemeRiver visualizations in Figures \ref{fig:BBSTopics} and \ref{fig:PRTopics}, the terms \emph{memory} and \emph{learning} are relatively prominent. This suggests that these may be the areas of interest, which demand the attention of future empirical studies in visualization. In particular, external memorization is an important merit of visualization \cite{Chen:2014:PIQ}, and there is a need to study how humans reduce their cognitive load for memorization through the use of visualization in addition to the current focus on how humans remember what is shown in visualization. Furthermore, learning is an important aspect of visualization as visualization can aid learning and learning can improve visualization skills. This line of enquiries may also lead to new scope of studies related to \emph{Developmental Psychology}.

While there are a small number of empirical studies in visualization designed to test some theories such as the Weber's law, the scale of using empirical studies to support theory development and model development is far smaller in comparison with that in psychology as illustrated by the terms \emph{model} and \emph{theory} in Figures \ref{fig:BBSTopics} and \ref{fig:PRTopics}, where there are many other terms, such as \emph{language}, \emph{information}, \emph{representation}, and \emph{social} that are also relevant to visualization but have not yet featured much in empirical studies in visualization.

\begin{table*}[t]
  \centering
  \caption{Many empirical studies have been conducted, by psychologists, statisticians, and visualization researchers, to examine humans' performance in estimating correlation using scatter plots and other visual representations. }
  \label{tab:Correlation}
  \scalebox{0.78}{
  	\begin{tabular}{@{}l@{\hspace{2mm}}c@{\hspace{2mm}}c@{\hspace{2mm}}c@{\hspace{2mm}}c@{\hspace{2mm}}c@{\hspace{2mm}}c@{\hspace{2mm}}c@{\hspace{2mm}}c@{\hspace{2mm}}c@{\hspace{2mm}}c@{\hspace{2mm}}c@{\hspace{2mm}}c@{}}
		& \textbf{Year} & \textbf{Venue} & \textbf{\#Authors} & \textbf{\#Experiment} & \textbf{\#Ind.Var.} & \textbf{Variations} & \textbf{\#Repeats} & \textbf{\#Iterations} & \textbf{\#Stimuli/P} & \textbf{\#Participants} & \textbf{Apparatus} & \textbf{Main Task} \\ \hhline{=============}
		Beach and Scopp~\cite{Beach:1966:PS} & 1966 & Psychology & 2 & 1 & 1 & A;5 $^a$ & 4 & & 20 & 32 & card & estimate $R$ \\ \hline
		Erlick~\cite{Erlick:1966:PS} & 1966 & Psychology & 1 & 1 & 2 & A:21$\times$2 & & & 50 & 10 & paper & \\ \hline
		Strahan and Hansen~\cite{Strahan:1978:APM} & 1978 & Psychology & 2 & 1 & 3 & A:13$\times$2;G:2 & & & 13 & $46+34$ & paper & \\ \hline
		Bobko and Karren~\cite{Bobko:1979:PP} & 1979 & Psychology & 2 & 1 & 3 & A:8;A:2;A:3 & & & 13 & 89 & questionnaire & estimate $R$ \\ \hline
		Lane et al.~\cite{Lane:1985:JEP} & 1985 & Psychology & 3 & 2 & & & & & & & & \\
		& & & & E1 & 5 & G:2;A:$2\times2\times2\times4$ & & & 40 & 39 & & estimate $R$ \\
		& & & & E2 & 5 & G:2;A:$2\times2\times2\times4$ & & & 40 & 40 & & estimate $R$ \\ \hline
		Lauer and Post~\cite{Lauer:1989:BIT} & 1989 & Psychology & 2 & 1 & 5 & A:$4\times3\times3\times2\times2$ & & & 144 & 27 & computer & estimate $R$ \\ \hline
		Collyer et al.~\cite{Collyer:1990:PMS} & 1990 & Psychology & 3 & 1 & 1 & A:4 & 4 & & 16 & 50 & computer & estimate $R$ \\ \hline
		Meyer and Shinar~\cite{Meyer:1992:HF} & 1992 & Psychology & 2 & 2 & & & & & & & & \\
		& & & & E1 & 4 & A$3\times3$;G:2;G:2 & & & 54 & 19+10 & paper & estimate $R$ \\
		& & & & E2 & 4 & A:$6\times3$;G:2;G:2 & & & 36 & 49+49 & paper & estimate $R$ \\ \hline
		Doherty et al.~\cite{Doherty:2007:PP} & 2007 & Psychology & 4 & 4 & & & & & & & & \\
		& & & & E1 & 1 & A:2 & & & 2 & 20 & paper & estimate $R$ \\
		& & & & E2 & 1 & A:4 & 25 & & 100 & 21 & paper & high/low \\
		& & & & E3 & 1 & A:4 & 25 & & 100 & 20 & paper & high/low \\
		& & & & E4 & 3 & A:$4\times2\times2$ & & & 1 $^b$ & 58 & paper & estimate $R$ \\ \hline
		Rensink~\cite{Rensink:2016:PBR} & 2017 & Psychology & 1 & 4 $^c$ & & & & & & & & \\
		& & & & E1 & 1 & A:10 & & \texttt{\char`\~}40 & & & computer & high/low \\
		& & & & E2 & 2 & A:$10\times2$ & & \texttt{\char`\~}40 & & & computer & high/low \\
		& & & & E3 & 2 & A:$10\times2$ & & \texttt{\char`\~}40 & & & computer & high/low \\
		& & & & E4 & 2 & A:$10\times2$ & & \texttt{\char`\~}40 & & & computer & high/low \\ \hhline{=============}
		Cleveland et al.~\cite{Cleveland:1982:S} & 1982 & Science & 3 & 3 & & & & & & & & \\
		& & & & E1 & 2 & A:$10\times4$ & 1,4 & & 19 & 74 & paper & estimate $R$ \\
		& & & & E2 & 1 & A:2 & & & 2 & 109 & projector & estimate $R$ \\
		& & & & E3 & 1 & A:2 & & & 2 & 32 & projector & estimate $R$ \\ \hhline{=============}
		Rensink and Baldridge~\cite{Rensink:2010:CGF} & 2010 & Visualization & 2 & 1 & & & & & & & & \\
		& & & & E1 & 1 & A/G:19 & & <50 & & 20 & computer & high/low \\
		& & & & E2 & 2 & A/G:$7\times2$ & & <50 & & 20 & computer & high/low \\ \hline
		Li et al.~\cite{Li:2010:IV} & 2010 & Visualization & 3 & 1 & 4 & A:$3\times7\times2\times2$ & 2 & & 168 & 25 & computer & estimate $R$ \\ \hline
		Harrison et al.~\cite{Harrison:TVCG:2014} & 2014 & Visualization & 4 & 2 & & & & & & & & \\
		& & & & E1 & 2 & G:6;A:2 & & <50 & <200 & 88 & crowd & high/low \\
		& & & & E2 & 3 & G:9;G:6;A:2 & & <50 & <200 & 1687 & crowd & high/low \\ \hline
		Kanjanabose et al.~\cite{Kanjanabose:CGF:2015} & 2015 & Visualization & 3 & 1 & 3 & A:$4\times3\times3$ & 2 & & 72 & 43 & computer & 4 tasks $^d$ \\ \hline
		Sher et al.~\cite{Sher:2017:CGF} & 2016 & Visualization & 4 & 1(6)$^e$ & & & & & 190 & 37 & & \\
		& & & & E1 & 1 & A:21 & >2 & & & & computer & estimate $R$ \\
		& & & & E2 & 2 & A:$3\times5$ & >2 & & & & computer & estimate $R$ \\
		& & & & E3 & 2 & A:$3\times3$ & >2 & & & & computer & estimate $R$ \\
		& & & & E4 & 1 & A:6 & >2 & & & & computer & estimate $R$ \\
		& & & & E5 & 1 & A:7 & >2 & & & & computer & estimate $R$ \\
		& & & & E6 & 2 & A:$4\times6$ & & & & & computer & estimate $R$ \\ \hhline{=============}
		Notes:\\
		\multicolumn{13}{l}{$a$. The letter ``A'' indicates that the variations were presented to all participants, while the letter ``G'' indicates that the variations were presented separately to different groups.}\\
		\multicolumn{13}{l}{$b$. Each participant received only 1 stimulus.}\\
		\multicolumn{13}{l}{$c$. The data resulting from the first experiment was used in the other three experiments as one of the two variations of the two-value variables.}\\
		\multicolumn{13}{l}{$d$. The four tasks are value retrieval, clustering, outlier detection, and change detection.}\\
		\multicolumn{13}{l}{$e$. The stimuli of six experiments were presented to participants in an integrated experiment to facilitate sharing of stimuli and alleviating cross-experiment confounding effects.}\\
	\end{tabular}
  }
  \vspace{-3mm}
\end{table*}

\section{Juxtaposed Case Studies}
\label{sec:CaseStudies}
The previous section juxtaposes the topic development in psychology and the empirical studies in visualization. With this broad context, this section focuses on two specific topics, for which we juxtapose the empirical studies published in psychology venues and those in visualization venues.

\subsection{Visual Estimation of Correlation}
Visually estimating correlation has been an interesting topic to psychologists, statisticians, and visualization researchers.
Table \ref{tab:Correlation} lists a number of empirical studies published mainly in psychology and visualization journals, including some major attributes of these studies.
The earlier studies on this topic in the 1960s, 1970s, 1980s, and 1990s typically involved apparatuses such as cards, papers and booklets, and projectors, while the use of computers started in the late 1980s and crowd-sourcing started in the current decade~\cite{BorgoMicallef:2018}.
Noticeably the introduction of computers as apparatuses enables more complicated study design, such as dynamic stimulus generation for iterative capture of participants' responses \cite{Rensink:2010:CGF,Harrison:TVCG:2014,Rensink:2016:PBR} and integrated multi-hypotheses experiment with stimuli sharing in \cite{Sher:2017:CGF}.
The crowd-based study conducted by Harrison et al.~\cite{Harrison:TVCG:2014} also demonstrated the feasibility of recruiting many more participants through the internet than any typical laboratory setting.
Possibly because of the programming skills available in the visualization community, the studies by Li et al.~\cite{Li:2010:IV}, Harrison et al.~\cite{Harrison:TVCG:2014},  Kanjanabose et al.~\cite{Kanjanabose:CGF:2015} enabled the investigation on this topic to be extended from visually estimating correlation using scatter plots to other visual representations for estimating correlation  and other visualization tasks using scatter plots.

The majority of the studies on this topic, in all publication venues, have identified that humans' estimation of the Pearson's product-moment correlation coefficient (PPMCC) do not have the same numerical accuracy and consistency as the PPMCC itself.
Some sounded an alarm about humans' sub-optimal inferences (e.g., \cite{Beach:1966:PS}), while others tried to model such displacement (e.g., \cite{Erlick:1966:PS,Cleveland:1982:S}).
Some 40 years ago, the psychologist authors of \cite{Bobko:1979:PP} suggested that ``\emph{examination of scatter plots may have many uses (cf. Tukey, 1977), although it is clear that calculation of \textbf{r} is not one of them.}'' Recently, visualization researchers built on the collective knowledge gained from the studies in Table \ref{tab:Correlation}, and started to ask the question about the benefits of scatter plots (e.g., \cite{Sher:2017:CGF}), and more broadly and deeply, the benefit of visualization in general \cite{Chen:2016:TVCG}.
All these enable us to appreciate the values of empirical studies such as those in Table \ref{tab:Correlation}, and motivate us to use empirical studies to help answer some fundamental questions in the field of visualization.  

\subsection{Color Perception and Colormapping}

Color perception has been a pervasive topic in psychology and visualization. The discipline of psychology has accumulated a very large collection of research papers on empirical studies and theoretical discourses derived from empirical studies. We provide below several examples of color research in psychology.

\begin{figure}
    \centering
    \includegraphics[height=45mm]{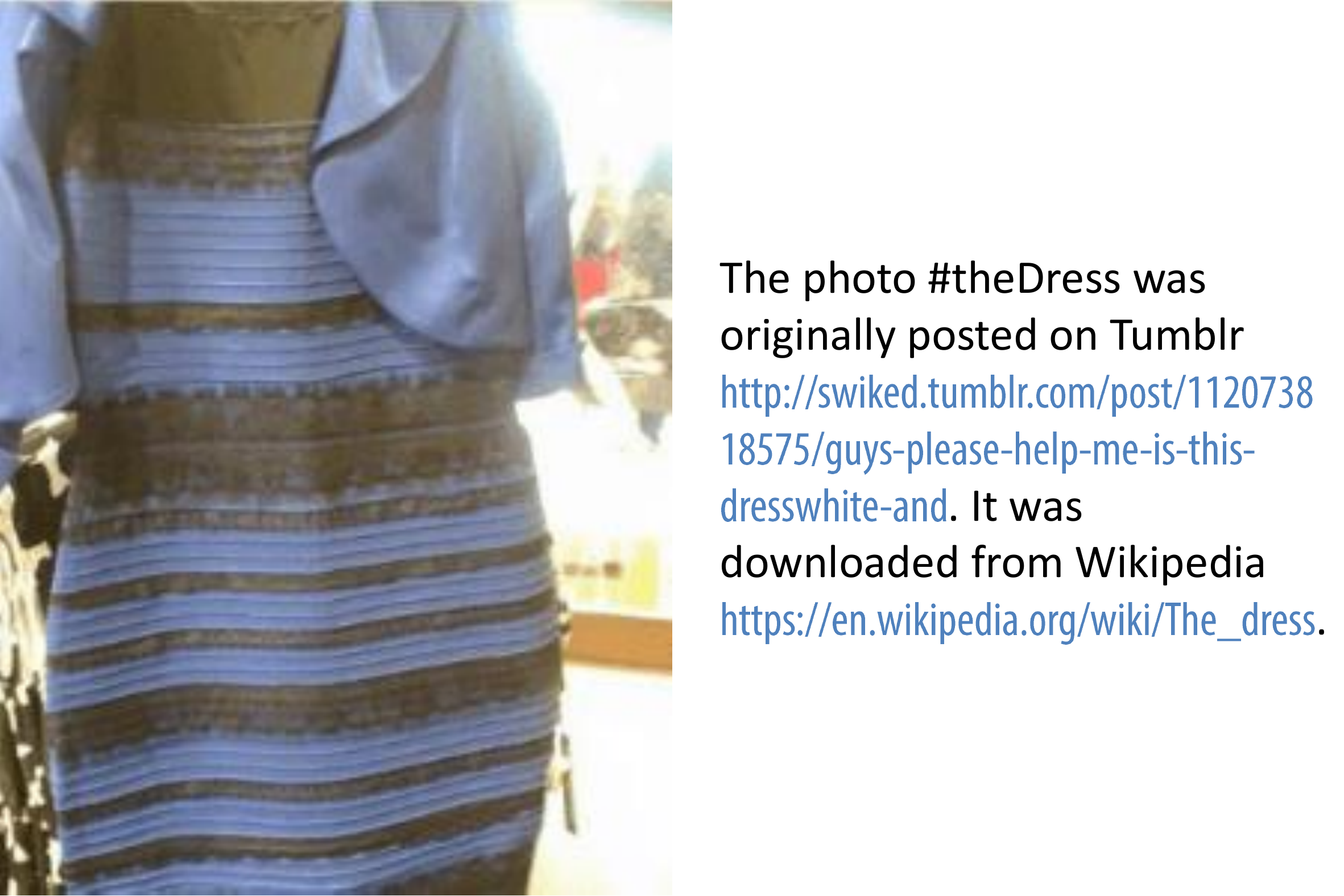}
    \caption{A photo, \#theDress, which first appeared on the social media service Tumblr, attracted a huge amount of online discussion. People perceived the colors of the dress differently --- many saw it in blue and lace black, while some saw white and gold.}
    \label{fig:theDress}
    \vspace{-5mm}
\end{figure}

\begin{enumerate}[noitemsep,nolistsep]
    \item Between late 1960s and early 1980s, Treisman and her colleagues reported a number of experiments for studying the interaction between colors and a few other visual channels (e.g., words \cite{Treisman:1969:N} and shapes \cite{Treisman:1977:book,Treisman:1977:PP}), which led to the proposal of a feature integration theory of attention \cite{Treisman:1980:CP}. Around that time, there were also many other publications in psychology, reporting experiments on colors and other visual channels as well as their preattentive properties, interactions, and integration (e.g., \cite{Shepard:1964:JMP,Williams:1967:AP,Burns:1978:PLM,Quinlan:1987:PP})  
    \item Colors were used in a number of experiments to study the effects of languages on cognition, e.g., color space in naming and memory \cite{Heider:1972:CP} (1972), color categories \cite{Davidoff:1999:N}, and preattentive color perception \cite{Thierry:2009:PNAS} (2009). The recent article by Zhong et al. mentioned some 20 references reporting empirical studies on this topic \cite{Zhong:2018:CS}.   
    %
    %
    \item Color perception was also a common topic shared by many branches of psychology, ranging from the leftmost branch in Figure \ref{fig:taxonomyH}, Biological Psychology, (e.g., \cite{Shapley:2002:CON}) to the rightmost branch in Figure \ref{fig:taxonomyH}, Comparative Psychology (e.g., \cite{Gouras:1979:JN}).
    %
    %
    \item Color constancy is a perceptual phenomenon that the colors on a surface appear to be constant despite measurable variations of intensity and spectrum  due to illumination and textures. Foster provided a substantial review on this topic, included many references \cite{Foster:2011:VR}. In 2015, an image, which was referred to as ``\#theDress'' online (Figure \ref{fig:theDress}), attracted a considerable amount of discussion in social media as different viewers appeared to perceive different colors of the dress. \#theDress also stimulated much scholarly discourse and some research activities among researchers in psychology. The recent review by Mart\'{i}n-Moro et al. included 17 references on this topic, including six empirical studies \cite{MartinMoro:2018:ASEO}, while Witzel and Gegenfurtner summarized the discussion on \#theDress in their review of the two closely related topics, color constancy and color categorization \cite{Witzel:2018:ARVS}.
\end{enumerate} 

In visualization, understanding color perception is vital to the design decisions on choosing visual channels and creating colormaps. Naturally visualization researchers have conducted many empirical studies on color perceptions and color mapping. Below are a number of empirical studies (in chronological order) that were published in visualization venues and collected during this survey.

\renewcommand{\labelenumi}{(\arabic{enumi})}
\begin{enumerate}[noitemsep,nolistsep]
    \item Ware conducted three experiments on color sequencing in colormaps \cite{Ware:1988:CGA}.
    \item Borgo et al. reported three experiments to study (i) the performance of five visualization tasks in spatio-temporal visualization using color pixel blocks, (ii) the effect of different numbers of color bands in a colormap, and (iii) the humans' capability of ``averaging'' in pixel-based visualization \cite{Borgo:2010:TVCG}. Their experiment (ii) confirmed Ware's finding about the merit of multi-band colormaps \cite{Ware:1988:CGA}.
    \item Haroz and Whitney reported three experiments to study the impact of visual feature type (color vs. motion), layout, and variety of visual elements on user performance \cite{Haroz:TVCG:2012}.
    \item  Griffin and Robinson compared the uses of colors and leader lines for highlighting visual objects depicted in coordinated views for geo-visualization \cite{Griffin:TVCG:2015}.
    \item Lin et al. conducted two experiments to compare the use of standard colormaps in visualization with the use of expert- or algorithm-selected colormaps with strong semantic association between colors and words \cite{Lin:CGF:2013}.
    \item Gramazio et al. reported an empirical study to examine the performance of color-based visual search tasks in three types of pixel grid layouts \cite{Gramazio:TVCG:2014}.
    \item Demiralp et al. reported an empirical study investigating the interaction between colors and a few other visual channels \cite{Demiralp:2014:TVCG}.
    \item Mittelst\"{a}dt and Keim conducted an experiment to study the impact of contrast effect on visualization tasks relying on color perception, and the means for alleviating the effect using personalized perception models \cite{Mittelstadt:CGF:2015}.
    \item Gramazio et al. reported an empirical study to evaluate a web-based tool for creating discriminable and aesthetically preferable categorical color palettes \cite{Gramazio:2017:TVCG}.
    \item Szafir reported three experiments to study the impact of mark types and sizes upon the perception of color differences in visualization \cite{Szafir:2018:TVCG}.
    \item Schloss et al. conducted an empirical study to examine the relationship between the semantic meaning associated to a sequential colormap and the ordering of the colors in the colormap \cite{Schloss:2019:TVCG}.
\end{enumerate}

Comparing the above two lists, we can easily observe the synergy between:
\begin{itemize}[noitemsep,nolistsep]
    \item (a) and (3), (4), (7) and (10);
    \item (b) and (5) and (11);
    \item (d) and (1), (2), (6), (8), and (9). 
\end{itemize}
The phenomenon exhibited by \#theDress in Figure \ref{fig:theDress} is directly related to many visualization tasks, especially those involving continuous colormaps and 3D visual objects.
Similarly to the case study in the previous section, if retrieving values from the colors of visual objects is not reliable, visualization researchers not only need to devise new guidelines, methods, and techniques to alleviate such problems, but also need to conduct more empirical studies that will help to answer the fundamental question ``what is really the benefit of visualization'' with unreliable visualization tasks for value retrieving. 

\section{Some Recent Developments in Psychology}
\label{sec:NewTrends}
Psychology is a continuing evolving discipline and new topics are emerging frequently.
In this section, we briefly describe three new developments and discuss their relevance to visualization. 

\subsection{Distributed Cognition Approaches in Cognitive Science}
\label{sec:D-Cognition}
While we classify Behavioural and Brain Sciences as a psychology journal in Section \ref{sec:TopicAnalysis}, it is also thought of as a journal in the interdisciplinary field of \emph{cognitive science}. Cognitive scientists seek converging evidence about the nature of cognition from multiple disciplines that speak to how information is processed, such as artificial intelligence, neuroscience, philosophy of mind, as well as psychology and social sciences.

A foundational principle in cognitive science is the concept of a \emph{cognitive architecture}, the structures of information processing that are architectural in the sense that they are invariant with regard to training, and experience. These are studied empirically in humans and simulated in computational cognitive architectures. For human cognitive architecture we see well-known neuroscience constraints such as trichromacy as determined by cone pigments and visual resolution limitations as determined by retinal receptor density. Other architectural limitations are defined by consistent limitations in human performance across tasks. These include attentional limitations, e.g., the number of spatial tokens (i.e., FINSTs) that parse complex visual scenes~\cite{Pylyshyn:2001:Cog}. These architectural constraints can be found in  the experimental psychology literature, however there are aspects of the cognitive architecture that relate specifically to interaction with dynamic and immersive visual environments that are not commonly studied by psychologists. Examples of these human/computer cognitive systems applications include: 

\begin{itemize}[noitemsep,nolistsep]
\item \emph{Smart seeing and projecting}~\cite{Kirsh:2010:AISoc}:  The argument for visualization often stems from our ability to see patterns in information graphics. This depends upon our mental models of the processes associated with that information as well as our ability to parse the artificial visual scene of the dashboard. A theory of the operating characteristics of \emph{smart seeing} visualization could be quite useful for creating and evaluating interactive methods. A related kind of expertise, \emph{projecting} is the ability of an expert to take into account what is represented in the visualization and to predict what will happen (or what should be done) next, then manipulating the information for a what-if analysis.
\item \emph{Enactive/complementary cognition}~\cite{Barsalou:2008:ARP}: Another way of integrating information technology and cognitive processes is when a dynamic environment generates information based on computational processes or changes in streaming data. Analysts must adjust their thinking and respond to the updated information in real time. Studies in the human factors literature document how changes in timing of the response to user actions can alter users' task performance strategy~\cite{Gray:2009:CHIWorkshop}. The ability of a theory to model human performance in dynamic environments would require it to be able to take the temporal coordination of cognitive processes and external events into account. Methods for doing this are still being developed, see \cite{Gray:2007:book, Kirsh:1994:CS, Lindstedt:2019:CP} for examples.
\item \emph{Multi-agent cognition and joint activity}~\cite{Kaastra:2014:Beliv}: The third D-Cog method studies coordination of action between multiple human and/or non-human agents, either through structured coordination protocols or as negotiated coordination. Two mechanisms can be used to enable negotiation: representation of the probable behaviours of an agent (e.g., a user behavioural model for an artificial intelligent agent) and cooperative signalling. Multi-human cognition is often studied using descriptive social science methods. A more focused approach comes from cognitive ethnography~\cite{Hutchins:1995:book}. A cognitivist research approach to multi-agent coordination examines human-human coordination as a Joint Activity~\cite{Bangerter:2003:CS}.  Joint activity models have been used to analyze coordinated activity in \emph{paired analysis} studies~\cite{Arias-Hernandez:2011:CS}. To do this an analysis task is proposed with roles given to two or more analysts. The roles require them to cooperate in accomplishing the analysis task in an interface environment~\cite{Kaastra:2014:Beliv}. Sessions are video captured and analyzed using Clark's theory. It is possible that this approach could also be extended to study of coordination with non-human agents.
\end{itemize}

\subsection{Mindfulness}
\label{sec:Mindfulness}
Third wave therapies such as mindfulness are becoming more popular in psychology. Though these are typically used in a health psychology context they also relate to cognition. Mindfulness can be defined as paying attention to the present moment, and in a non-judgmental way~\cite{Kabat-Zinn:2003:CPSP}. Mindfulness meditation is thought to promote cognitive flexibility which may be useful in visualization tasks. For example, after mindfulness training individuals encountered less stroop task interference (a measure of automatic thinking) and performed better in the $d2$ concentration and endurance test. The $d2$ task requires the individual to visually discriminate targets from visually similar non-targets. So, mindfulness may have increased the visual attention of participants~\cite{Moore:2009:CC}.

In many visual analytics applications, analysts may encounter a variety of psychological conditions that may impact on the performance of visual analytics tasks. Research on mindfulness in the context of visualization and visual analytics may provide new means to address such conditions.

\subsection{Cognitive Neuroscience}
\label{sec:C-Neuroscience}
\emph{Cognitive neuroscience} \cite{Andersen:1992:book,Ochsner:2017:book} is an interdisciplinary field connecting neuroscience with psychology. It is a topic category under the leftmost branch in Figure \ref{fig:taxonomyH}, Biological Psychology, which is concerned with the biological processes and aspects that underlie human behaviors and cognition.
Cognitive neuroscience focuses on the neural connections in the brain, their formation and transformation, their functions and their controls, and their impact on various cognitive processes (cf. the 7th variables in Figure \ref{fig:taxonomyH}).

Over the past four decades, the rapid advancement of new brain mapping technologies (e.g., fMRI and PET) has enabled cognitive scientists to observe brain activities at a more detailed spatiotemporal scale than ever before. These technologies have been used to study human vision systems (e.g., \cite{Courtney:1997:CON,Hickey:2015:N,Schwartz:2005:CC}), and visualization-related cognitive functions such as memory and reasoning (e.g., \cite{Carpenter:2000:CON,Durning:2015:MT}). The applications of visualization and visual analytics have not been at the same scale as other imaging modalities (e.g., CT, MRI, DTO, etc.), though there were some reports of such applications (e.g., \cite{Katwal:2013:TBE,Kasabov:2017:TNNLS}).
There is a huge potential for developing advanced visualization and visual analytics techniques in supporting functional neuroimaging and hence cognitive neuroscience.

Meanwhile, more advanced and effective analysis of functional neuroimaging data will provide the field of visualization with more opportunities to study visualization phenomena using functional neuroimaging.

\section{Challenges and Opportunities}
\label{sec:Challenges}
There are many branches of \emph{Applied Psychology}, some of which are shown in Figure \ref{fig:taxonomyH}. One has to ask that ``is there a room for \emph{Visualization Psychology}?'' The authors of this survey believe that the visualization community should work with colleagues in psychology to establish such a branch. We hope that this survey is an early step towards this long term goal.

There will be many challenges along the route to the establishment of a new branch of \emph{Applied Psychology}. These may include:
\begin{itemize}
\item Many research students in visualization may need some persuasion to take on empirical studies as their thesis topics.
\item Many academic supervisors may feel uncomfortable to start a new line of scientific investigation.
\item The perception about the relatively lower acceptance rate for papers in the category of ``Evaluation'' or ``Empirical Studies''.
\end{itemize}

Meanwhile, visualization provides a unique window on the human mind, while playing an indispensable role in data science. This surveys shows that the visualization community is not only capable of carrying out empirical studies to test some visual designs or visualization systems as part of a software engineering workflows but also capable of attempting the more ambitious goal of empirical studies, that is, to make new discoveries about how and why visualization works in some conditions and not in others, and to inform and verify proposed theories advances.

Most of us agree that in some circumstances, visualization is more effective and/or efficient than viewing data in numerical, textual, or tabular forms, and than being simply informed by a computer about the decision. When visualization works in these circumstances, there must be some merits in perception and cognition. Hence any causal factors that make visualization work may potentially be the causal factors that make perception and cognition work. Therefore, visualization researchers are in the right place at the right time to look for these causal factors.

In summary, while the field of visualization can learn a huge amount from psychology in terms of research findings and research methodologies, there is a need to develop an interdisciplinary subject, bringing together the discipline of psychology and the field of visualization more closely. While visualization can be a significant application area of psychology, visualization researchers can also provide advanced computing technologies to support the design of empirical studies and the analysis of captured empirical data.
While there is a continuing need to conduct usability studies for evaluating visual designs, visualization techniques, and visualization systems, there is profound need to design innovative empirical studies for understanding complex phenomena in visualization and for informing the development of the foundation of visualization.


\bibliographystyle{IEEEtran}
\bibliography{updated-references,new-references}


%

\begin{IEEEbiography}[{\includegraphics[width=1in,height=1.25in,clip,keepaspectratio]{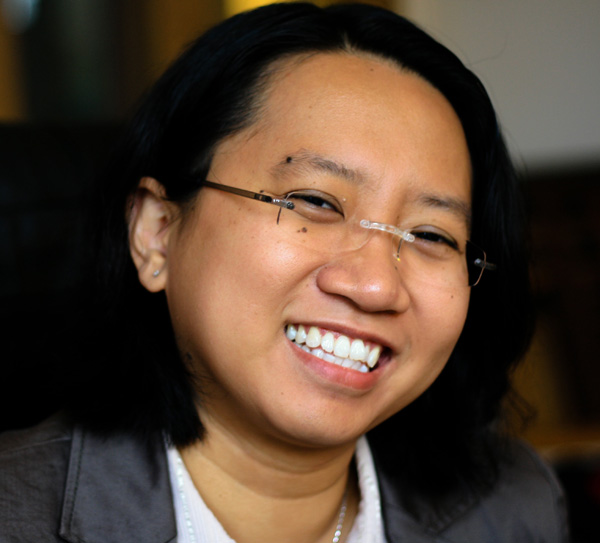}}]{Alfie Abdul-Rahman.}
Alfie Abdul-Rahman is a Lecturer in Computer Science at King's College London. She received her PhD from Swansea University in computer science. Before joining King's College London, she was a Research Associate at the University of Oxford e-Research Centre. She worked as a Research Engineer in HP Labs Bristol on document engineering, and then as a software developer in London, working on multi-format publishing. Her research interests include visualization, computer graphics, and human-computer interaction.
\end{IEEEbiography}

\vfill

\begin{IEEEbiography}[{\includegraphics[width=1in,height=1.25in,clip,keepaspectratio]{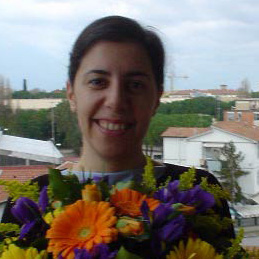}}]
{Rita Borgo.}
Dr Rita Borgo is a Senior Lecturer in the Informatics Department at King’s College (KCL) and currently Head of the Human Centred Computing research group. Her main research interests lie in the areas of Information Visualization and Visual Analytics with particular focus on the role of Human Factors in Visualization. Her research has followed an ambitious program of developing new data visualization techniques for interactive rendering and manipulation of large multi-dimensional and multivariate datasets. Novel in all aspects of the research is the aim at providing solutions that involve human in the loop of intelligent reasoning while reducing the burden of inspection of large complex data. Her research has been awarded supports from Royal Society, EPSRC and EU. She is currently championing the newly created Urban Living hub at KCL and works in close collaboration with the Centre for Urban Science and Progress (CUSP) – London to increase impact of visualization within urban related challenges.
\end{IEEEbiography}

\vfill

\begin{IEEEbiography}[{\includegraphics[width=1in,height=1.25in,clip,keepaspectratio]{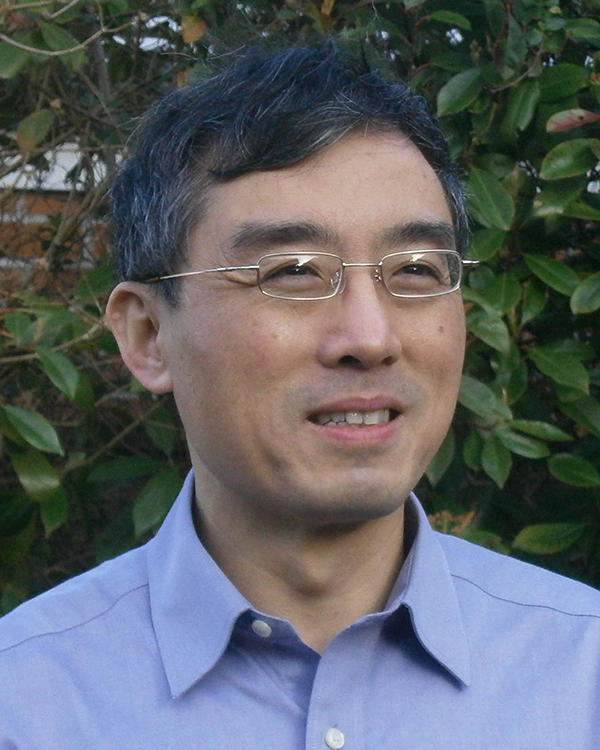}}]{Min Chen.}
Min Chen received his PhD degree from University of Wales in 1991. He is currently a professor of scientific visualization at Oxford University and a fellow of Pembroke College. Before joining Oxford, he held research and faculty positions at Swansea University. His research interests include visualization, computer graphics and human-computer interaction. His services to the research community include papers co-chair of IEEE Visualization 2007 and 2008, IEEE VAST 2014 and 2015, and Eurographics 2011; co-chair of Volume Graphics 1999 and 2006, and EuroVis 2014; associate editor-in-chief of IEEE TVCG; editor-in-chief of Computer Graphics Forum; and co-director of Wales Research Institute of Visual Computing. He is a fellow of BCS, EG and LSW.
\end{IEEEbiography}

\vfill

\begin{IEEEbiography}[{\includegraphics[width=1in,height=1.25in,clip,keepaspectratio]{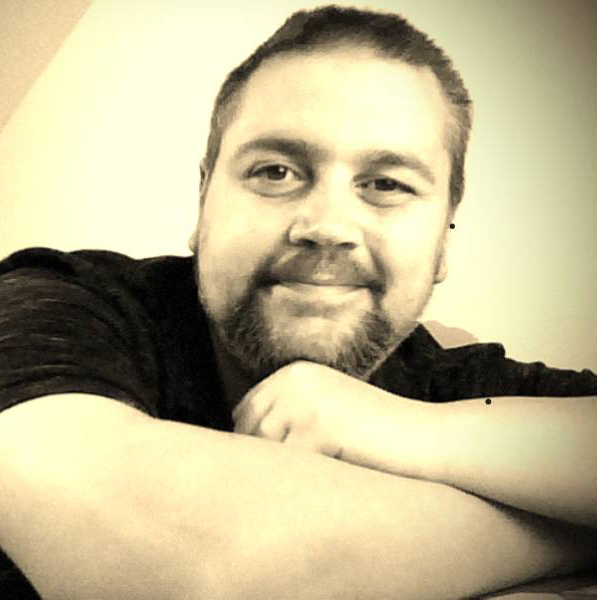}}]{Darren J. Edwards.}
Dr. Darren J. Edwards is a cognitive, health, and experimental researcher at Swansea University (Senior Lecturer). His research has explored information theories of cognition, particularly in the area of information reduction, simplicity, and categorization tasks. Darren has worked with computer science academics for several years, applying theories of psychology and cognition to areas such as data visualization and human computer interaction.
\end{IEEEbiography}

\vfill

\begin{IEEEbiography}[{\includegraphics[width=1in,height=1.25in,clip,keepaspectratio]{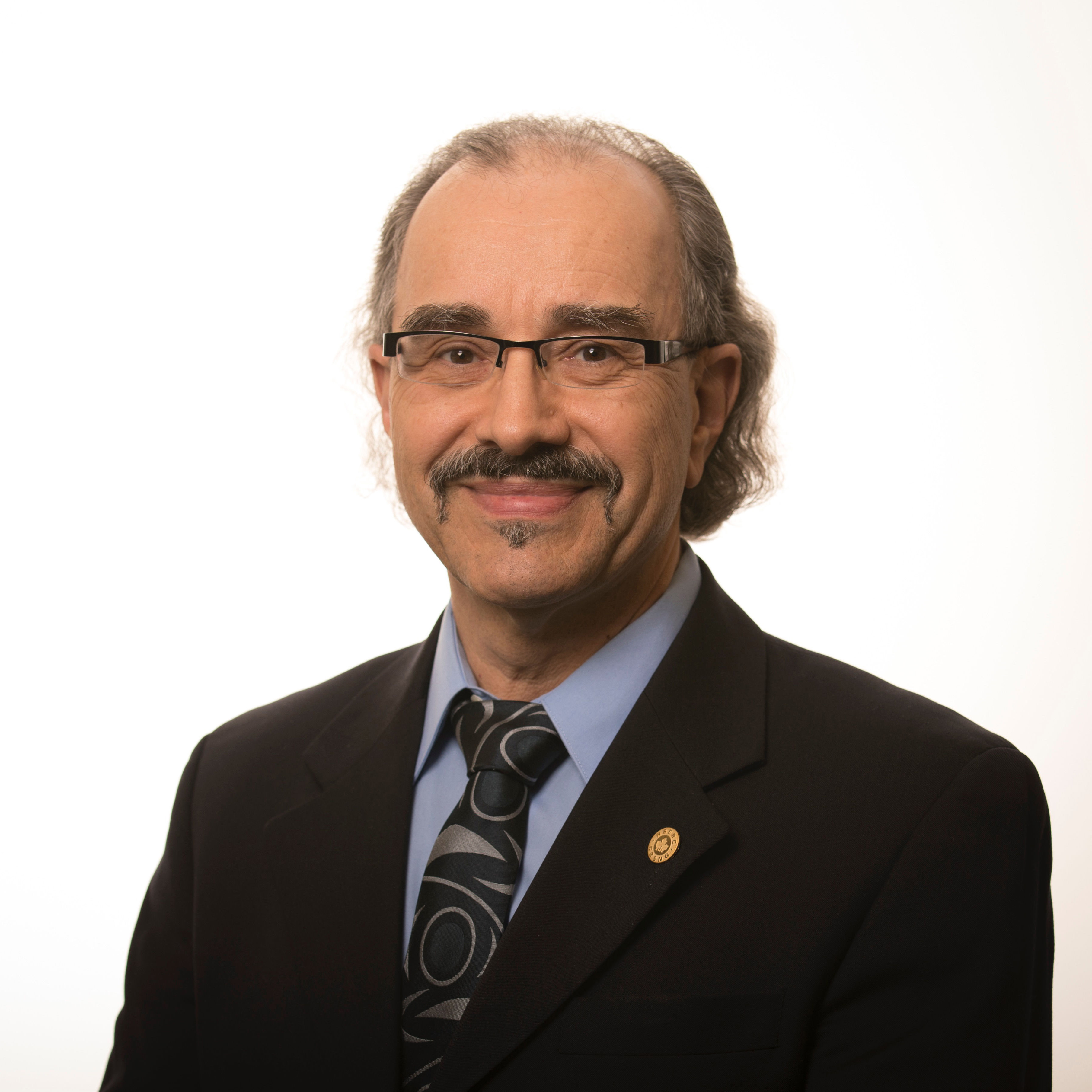}}]{Brian Fisher.}
Brian Fisher is a Professor in the School of Interactive Arts and Technology (SIAT) at Simon Fraser University (SFU) and Affiliate Professor in Computer Science at the University of British Columbia (UBC).  Earlier he was was Associate Director for the Media And Graphics Interdisciplinary Centre (MAGIC) and Adjunct Professor in Psychology, Computer Science, and Strategy and Business Economics at UBC. His research takes a cognitive science perspective on the design and evaluation of technology to support human understanding, decision-making, and collaboration in applications that include personalized health, aircraft safety analysis, and command, control, and interoperability for emergency management. Brian is a Fellow of the Psychonomics Society, and serves on the Natural Sciences and Engineering Research Council of Canada, the IEEE VAST Steering and VIS Executive Committees.
\end{IEEEbiography}

\end{document}